\documentclass[a4paper,12pt]{article}
\usepackage[a4paper,text={16.8cm,22.4cm}]{geometry}
\usepackage{amsmath,amssymb,bm,cite,psfrag,graphicx}
\usepackage{booktabs}
\usepackage[normalem]{ulem}

\usepackage{color}

\allowdisplaybreaks 
\renewcommand{\arraystretch}{2}

\newcommand{\pTveto}{p_T^{\hspace{-0.3mm}\rm veto}\hspace{-0.3mm}}

\begin{document}

\begin{titlepage}

\begin{flushright}
MITP/13-37\\
June 28, 2013
\end{flushright}
 
\vspace{0.3cm}
\begin{center}
\Large\bf\boldmath
Factorization and N$^3$LL$_{\rm p}+$NNLO Predictions for the Higgs Cross Section with a Jet Veto
\end{center}

\vspace{0.3cm}
\begin{center}
Thomas Becher$^a$, Matthias Neubert$^b$ and Lorena Rothen$^a$ \\
\vspace{0.4cm}
{\sl ${}^a$\,Albert Einstein Center for Fundamental Physics\\
Institut f\"ur Theoretische Physik, Universit\"at Bern\\
Sidlerstrasse 5, CH--3012 Bern, Switzerland\\[0.4cm]
${}^b$\,PRISMA Cluster of Excellence \& Mainz Institut for Theoretical Physics\\
Johannes Gutenberg University, D-55099 Mainz, Germany}
\end{center}

\vspace{0.5cm}
\begin{abstract}
\vspace{0.2cm}
\noindent 
We have recently derived a factorization formula for the Higgs-boson production cross section in the presence of a jet veto, which allows for a systematic resummation of large Sudakov logarithms of the form $\alpha_s^n\ln^m(\pTveto/m_H)$, along with the large virtual corrections known to affect also the total cross section. Here we determine the ingredients entering this formula at two-loop accuracy. Specifically, we compute the dependence on the jet-radius parameter $R$, which is encoded in the two-loop coefficient of the collinear anomaly, by means of a direct, fully analytic calculation in the framework of soft-collinear effective theory. We confirm the result obtained by Banfi et al.\ from a related calculation in QCD, and demonstrate that factorization-breaking, soft-collinear mixing effects do not arise at leading power in $\pTveto/m_H$, even for $R={\cal O}(1)$. In addition, we extract the two-loop collinear beam functions numerically. We present detailed numerical predictions for the jet-veto cross section with partial next-to-next-to-next-to-leading logarithmic accuracy, matched to the next-to-next-to-leading order cross section in fixed-order perturbation theory. The only missing ingredients at this level of accuracy are the three-loop anomaly coefficient and the four-loop cusp anomalous dimension, whose numerical effects we estimate to be small.
\end{abstract}
\vfil

\end{titlepage}

\section{Introduction}

With firm evidence for a Higgs boson with a mass around $m_H=125$\,GeV, the primary focus of particle physics has now shifted to the study of the properties of this new particle, in particular of its couplings. An important channel in this context is Higgs-boson production with subsequent decay into a $W^+ W^-$ pair, for which both ATLAS and CMS have recently reported $4 \sigma$ evidence \cite{ATLAS:2013wla,CMS:bxa}. With a branching ratio of about 22\%, this is the second largest decay channel of the Higgs boson. Because of the missing energy in the final state, the $W^+ W^-$ channel is not particularly well suited for a Higgs mass measurement, but it offers the possibility for a precise Higgs coupling measurement and spin studies. A challenge is posed by the large background from $t\bar t$ production, which, after the top-quarks decay, results in a $W^+ W^-$ pair in association with two $b$-quark jets. This background is significantly reduced by rejecting events containing jets with transverse momentum above a certain threshold $\pTveto$, which is chosen around 25--30\,GeV in current experimental analyses. 

Imposing such a jet veto enhances the higher-order QCD corrections to the Higgs-boson production cross section by Sudakov logarithms of the form $\alpha_s^n\ln^m(\pTveto/m_H)$, with $m\le 2n$. One might argue that these logarithms are not particularly large for the relevant values of $\pTveto$; however, even the total Higgs production rate suffers from large corrections, and additional enhancements can easily lead to unreliable predictions. For the fixed-order predictions of the production cross section with a jet veto, it was observed that there is a numerical cancellation between the negative corrections from Sudakov logarithms and the large positive virtual corrections to the total rate, which leads to artificially small scale uncertainties \cite{Anastasiou:2007mz}. To avoid this cancellation and get a more reliable estimate of the theoretical uncertainties, it was subsequently proposed to add in quadrature the scale uncertainties of the total cross section and the cross section with one or more jets in the final state, which leads to an uncertainty of 17\% on the Higgs production cross section with a jet veto \cite{Stewart:2011cf}. This uncertainty is about twice as large as the experimental systematic errors and of the same size as the current statistical uncertainty. To make full use of the coming LHC data, the theoretical uncertainty should thus be reduced significantly.

There has been a lot of progress in the theoretical description of the Higgs-boson production rate with a jet veto over the past year, starting with the work \cite{Banfi:2012yh}, where it was shown that the Sudakov logarithms associated with the jet veto can be resummed at next-to-leading logarithmic (NLL) order. Subsequently, we have derived an all-order factorization theorem \cite{Becher:2012qa} using soft-collinear effective theory (SCET) \cite{Bauer:2000yr,Bauer:2001yt,Bauer:2002nz,Beneke:2002ph}, which allows for resummation to any desired accuracy, given the necessary perturbative input. We have also explicitly carried out the resummation at NNLL order. One of the necessary ingredients at this level of accuracy is the two-loop coefficient $d_2^{\rm veto}(R)$ of the so-called collinear anomaly \cite{Becher:2012qa,Becher:2010tm}, which we had extracted from partial NNLL results of \cite{Banfi:2012yh} under the assumption that these results remained valid in the limit where the jet-radius parameter $R$ is taken to infinity. It was subsequently shown by Banfi et al.\ that this assumption does not hold \cite{Banfi:2012jm}. The limits $m_H\to\infty$ and $R\to\infty$ do not commute, and taking the large-$R$ limit naively one misses an $R$-independent term in $d_2^{\rm veto}(R)$. After correcting the value of the two-loop coefficient accordingly, there is full agreement between the NNLL results presented in \cite{Becher:2012qa} and \cite{Banfi:2012jm}. 

The validity of factorization formula put forward in \cite{Becher:2012qa} was questioned by the authors of \cite{Tackmann:2012bt}, who claimed that this formula breaks down unless the jet radius $R$ is assumed to be parametrically small, such that $R\sim\pTveto/m_H\ll 1$. However, for small $R$ the perturbative corrections to the cross section are enhanced by logarithms of the jet radius, and these logarithms cannot be resummed by means of the factorization formula obtained in \cite{Becher:2012qa}. Reference \cite{Tackmann:2012bt} concluded that one is ``stuck between a rock and a hard place'', because one would either face factorization-breaking corrections or large unresummed logarithms of $R$. However, immediately after the paper \cite{Tackmann:2012bt} appeared a NNLL resummation formula was published in \cite{Banfi:2012jm}, and it was verified numerically that it correctly predicts the relevant logarithms up to ${\cal O}(\alpha_s^3)$, even for $R\sim 1$. To NNLL accuracy, our factorization formula precisely matches the result of \cite{Banfi:2012jm}.

The purpose of the present paper is three-fold. First, we will compute the two-loop anomaly coefficient $d_2^{\rm veto}(R)$ directly within the SCET framework. We find complete agreement with the QCD result of \cite{Banfi:2012yh}, which demonstrates explicitly and analytically that our factorization formula, which does not include soft-collinear mixing terms, is correct up to NNLL order. We provide a completely analytic result for the expansion of the two-loop anomaly coefficient in $R$, whose $R$-independent piece was only obtained in numerical form in previous papers \cite{Banfi:2012yh}. Secondly, we will show that the soft-collinear mixing contributions obtained in \cite{Tackmann:2012bt}  are absent if one ensures that the computation is done in such a way that there is no double counting among the different momentum regions in the effective theory. This double counting is avoided from the beginning if loop and phase-space integrals in the effective theory are properly expanded in the different momentum regions. If the jet measure is left unexpanded, as was done in \cite{Tackmann:2012bt}, then non-zero soft-collinear mixing contributions can arise in individual integrals, but they cancel if the necessary subtractions are performed to remove the soft-collinear overlap regions from the integrals. We discuss these issues in detail, give arguments that factorization breaking will also not arise at higher logarithmic accuracy, and conclude that all the available evidence indicates that the factorization theorem proposed in \cite{Becher:2012qa} is valid to all orders. Thirdly, we present an updated and improved phenomenological analysis of the Higgs-boson production cross section with a jet veto. Since the two-loop anomaly coefficient turns out to be numerically large for the values of $R$ used by the experimental collaborations, the predictions obtained at NNLL order still suffer from significant scale uncertainties. We show that all the ingredients required to increase the accuracy to N$^3$LL order are either already known or can be extracted numerically, except for the three-loop anomaly coefficient and the four-loop cusp anomalous dimension. We estimate the effect of these missing coefficients and find that they only have a small numerical impact on the results. We thus obtain predictions with N$^3$LL$_{\rm p}$ accuracy, where the subscript ``p'' (for ``partial'') indicates that two of the ingredients for a complete N$^3$LL calculation are yet unknown. We also include power-suppressed terms by matching our results to the NNLO fixed-order cross section, finding that these power corrections are numerically small. This indicates that the expansion about small $\pTveto$ is well behaved at the experimentally relevant values of the jet-veto scale.

Our paper is organized as follows: We first review in Section~\ref{sec:2} the factorization theorem for the cross section with a jet veto and collect the necessary perturbative ingredients. In Section~\ref{sec:multipole}, we then discuss the clustering of particles in different momentum regions and show that factorization-breaking terms are absent. After this general discussion, we present the explicit calculation of the two-loop anomaly coefficient $d_2^{\rm veto}(R)$ in Section~\ref{sec:2loop}. The numerical extraction of the two-loop beam functions and the fixed-order matching are discussed in Section~\ref{sec:beamfuns}. With these ingredients at hand, we present in Section~\ref{sec:numerics} our numerical results for the jet-veto cross section for Higgs production at the LHC. Our conclusions are summarized in Section~\ref{sec:concl}. In the Appendix, we give some details on the analytic calculation of the two-loop anomaly coefficient as an expansion in the jet-radius parameter $R$.

\section{Factorization theorem for the jet-veto cross section}
\label{sec:2}

Using arguments based on SCET, we have shown in \cite{Becher:2012qa} that the Higgs-boson production cross section defined with a jet veto $p_T^{\rm jet}<\pTveto$ can be factorized, to all orders in perturbation theory and at leading power in the small ratio $\pTveto/m_H$, in a way that separates the short-distance scales $m_t$ and $m_H$ from the scale $\pTveto$ of the jet veto. We work with the usual class of sequential recombination jet algorithms \cite{Salam:2009jx}, with distance measure
\begin{equation}\label{algo}
   d_{ij} = \mbox{min}(p_{Ti}^n,p_{Tj}^n)\,
    \frac{\sqrt{\Delta y_{ij}^2+\Delta\phi_{ij}^2}}{R} \,, \qquad
   d_{iB} = p_{Ti}^n \,,
\end{equation}
where $n=1$ corresponds to the $k_T$ algorithm \cite{Catani:1993hr,Ellis:1993tq}, $n=0$ to the Cambridge-Aachen algorithm \cite{Dokshitzer:1997in,Wobisch:1998wt}, and $n=-1$ to the anti-$k_T$ algorithm \cite{Cacciari:2008gp}. The two particles with the smallest distance are combined into a new ``particle'', whose momentum is the sum of the momenta of the parent particles. If the smallest distance is $d_{iB}$, then particle $i$ is considered a jet and removed from the list. The procedure is iterated until all particles are grouped into jets, i.e., the algorithm is inclusive. In the following, the jet-radius parameter is assumed to obey the inequalities 
\begin{equation}\label{R1}
   \frac{\pTveto}{m_H}\ll R\ll \ln\frac{m_H}{\pTveto} \,,
\end{equation}
and we work in the limit where $\lambda=\pTveto/m_H$ is a small expansion parameter. Then these inequalities are satisfied as long as $R$ is treated as an ${\cal O}(1)$ number, independent of $\lambda$. For too small values of $R$ (meaning $R\sim\lambda$ or smaller), large logarithms $\ln^n\!R$ arise, which would require a special treatment. These ``clustering logarithms'' have a complicated structure in higher orders \cite{Kelley:2012kj,Kelley:2012zs}, and it is currently not understood how to resum them. For too large $R$ (meaning $R\sim\ln(1/\lambda)$ or larger), on the other hand, the factorization formula breaks down. 

The factorization formula is obtained by factorizing the contributions of hard, collinear, anti-collinear, and soft modes in SCET. Denoting by $y$ the rapidity of the Higgs boson in the proton-proton center-of-mass frame, one first derives the preliminary result
\begin{equation}\label{dsigdy}
   \frac{d\sigma(\pTveto)}{dy}
   = \sigma_0(\mu)\,C_t^2(m_t^2,\mu) \left| C_S(-m_H^2,\mu) \right|^2 
    \left[ {\cal B}_c(\xi_1,\pTveto,\mu)\,{\cal B}_{\bar{c}}(\xi_2,\pTveto,\mu)\,
    {\cal S}(\pTveto,\mu) \right]_{q^2=m_H^2} ,
\end{equation}
where $\xi_{1,2}=(m_H/\sqrt{s})\,e^{\pm y}$ and
\begin{equation}
   \sigma_0(\mu) = \frac{m_H^2\,\alpha_s^2(\mu)}{72\pi (N_c^2-1) s v^2} \,.
\end{equation}
The Wilson coefficient $C_t=1+{\cal O}(\alpha_s)$ arises when one approximates the fermion-loop contribution to the gluon fusion amplitude by an effective, local $Hgg$ operator, as is routinely done in calculations of the Higgs-boson production amplitude. The hard matching coefficient $C_S=1+{\cal O}(\alpha_s)$ appears when the scalar two-gluon operator is matched onto a corresponding operator in SCET \cite{Ahrens:2008nc}. Both coefficients are known to three-loop order in perturbation theory, but for our purposes we only need the two-loop expressions derived in \cite{Kramer:1996iq,Chetyrkin:1997iv} and \cite{Harlander:2000mg,Ahrens:2008nc}, respectively. The resulting expressions can also be found in eqs.~(12) and (17) of \cite{Ahrens:2008nc}.

The emissions of (anti-)collinear and soft gluons, which are then grouped into jets according to the jet algorithm, are accounted for by the beam functions ${\cal B}_c$, ${\cal B}_{\bar c}$ and the soft function ${\cal S}$ in the factorization theorem (\ref{dsigdy}). Besides the veto scale, these functions also depend on the jet definition and in particular on the jet-radius parameter $R$. This dependence is suppressed in our notation. The collinear matrix element relevant for Higgs production reads \cite{Becher:2012qa}
\begin{equation}\label{collfun}
\begin{aligned}
   {\cal B}_{c,g}(z,\pTveto,\mu)
   &= - \frac{z\,\bar n\cdot p}{2\pi} \int dt\,e^{-izt\bar n\cdot p}  
    \sum \hspace{-0.8cm}\int\limits_{X_c,\,{\rm reg.}} \!\!
    {\cal M}_{\rm veto}(\pTveto,R,\{ \underline{p_c} \}) \\[-1mm]
   &\quad\times \langle P(p)|\,{\cal A}_{c\perp}^{\mu,a}(t\bar n)\,|X_c\rangle\,
    \langle X_c|\,{\cal A}_{c\perp\mu}^a(0)\,|P(p)\rangle \,,
\end{aligned}
\end{equation}
where ${\cal A}_{c\perp}$ denotes the gauge-invariant collinear gluon field in SCET. The matrix element in the second line is exactly the same as that entering the definition of the standard parton distribution function (PDF) for the gluon. The only difference is that the sum over intermediate states in (\ref{collfun}) is constrained by the jet veto, whose effect is encoded in a ``measurement function'' ${\cal M}_{\rm veto}$, which depends on the momenta  $\{ \underline{p_c} \}$ of the particles in the final state. Likewise, the soft function is defined as
\begin{equation}\label{softfun}
   {\cal S}(\pTveto,\mu)
   = \frac{1}{d_R} \,\sum \hspace{-0.8cm} \int\limits_{X_c,\,{\rm reg.}} \!\!
    {\cal M}_{\rm veto}(\pTveto,R,\{ \underline{p_s} \}) \langle\,0\,|\,
    \big( S_n^\dagger S_{\bar n} \big)^{ab}(0)\,|X_s\rangle\,
    \langle X_s|\,\big( S_{\bar n}^\dagger S_n \big)^{ba}(0)\,|0\rangle \,,
\end{equation}
with $d_R=N_c^2-1$. It involves Wilson lines of soft gluon fields in the adjoint representation, integrated along the beam directions $n$ and $\bar n$. 

Like in the case of the transverse-position dependent PDFs studied in \cite{Becher:2010tm}, the presence of a measurement function probing parton transverse momenta leads to additional light-cone (or rapidity) divergences, which are not regularized in dimensional regularization. The sums over collinear states $X_c$ in (\ref{collfun}) and soft states $X_s$ in (\ref{softfun}) are therefore regularized analytically. To this end, we use the phase-space regularization prescription of \cite{Becher:2011dz}, which amounts to replacing the usual phase-space measure by 
\begin{equation}\label{analytreg}
   \int\!d^dk\,\delta(k^2)\,\theta(k^0) ~ \to ~
   \int\!d^dk \left( \frac{\nu}{k_+} \right)^\alpha \delta(k^2)\,\theta(k^0)
   =\frac{1}{2} \int\!dy\!\int\!d^{d-2}k_\perp \left( \frac{\nu}{k_T} \right)^\alpha e^{-\alpha \,y} \,,
\end{equation}
where $y=\frac12\ln(k_+/k_-)=\ln(k_+/k_T)$ and $k_T=|\vec k_\perp|$. The regularization softens the light-cone singularities arising in the evaluation of the matrix elements. It introduces a new scale $\nu$, which plays an analogous role to the scale $\mu$ entering in dimensional regularization.

Once the light-cone singularities in the (anti-)collinear and soft functions have been regularized, they show up as poles in the analytic regulator $\alpha$, which cancel in the product of the three matrix elements in (\ref{dsigdy}). However, after the cancellation large logarithms of the scale ratio $m_H/\pTveto$ arise, which need to be resummed to all orders in perturbation theory. This effect has been called the ``collinear factorization anomaly'' \cite{Becher:2010tm}. The resummation of the anomalous logarithms can be accomplished by means of solving simple differential equations, which express the fact that the product of the three functions must be regularization independent \cite{Becher:2012qa}. One finds that
\begin{equation}\label{colanom}
\begin{aligned}
   & \left[ {\cal B}_c(\xi_1,\pTveto,\mu)\,{\cal B}_{\bar{c}}(\xi_2,\pTveto,\mu)\,
    {\cal S}(\pTveto,\mu) \right]_{q^2=m_H^2} \\
   &= \left( \frac{m_H}{\pTveto} \right)^{-2F_{gg}(\pTveto,\mu)} e^{2h_A(\pTveto,\mu)}\,
    \bar B_g(\xi_1,\pTveto)\,\bar B_g(\xi_2,\pTveto) \,,
\end{aligned}
\end{equation}
where the anomalous dependence on the hard scale $m_H$ is now explicit. Compared with \cite{Becher:2012qa}, we have extracted a factor $e^{h_A(\pTveto,\mu)}$ from each collinear function, which is chosen such that the remaining function $\bar B_g(\xi,\pTveto)$ is renormalization-group (RG) invariant. We have also absorbed the square root of the soft function into the collinear matrix elements. (In the regularization scheme adopted here, ${\cal S}(\pTveto,\mu)=1$ to all orders in perturbation theory, so this last step is trivial.) The exponents $F_{gg}$ and $h_A$ obey the RG equations \cite{Becher:2011xn,Becher:2010tm}
\begin{equation}\label{evol}
\begin{aligned}
   \frac{d}{d\ln\mu}\,F_{gg}(\pTveto,\mu) 
   &= 2\Gamma_{\rm cusp}^A(\mu) \,, \\
   \frac{d}{d\ln\mu}\,h_A(\pTveto,\mu) 
   &= 2\Gamma_{\rm cusp}^A(\mu)\,\ln\frac{\mu}{\pTveto} - 2\gamma^g(\mu) \,,
\end{aligned}
\end{equation}
where without loss of generality we can impose the normalization condition $h_A(\pTveto,\pTveto)=0$. In (\ref{evol}), $\Gamma_{\rm cusp}^A$ is the cusp anomalous dimension in the adjoint representation, and $\gamma^g$ denotes the anomalous dimension of the collinear gluon field as defined in \cite{Becher:2009qa}. For our analysis we require the three-loop expression for the anomaly exponent $F_{gg}$ and the two-loop result for $h_A$. Solving the evolution equations (\ref{evol}), we obtain
\begin{eqnarray}\label{Fhexpansions}
   F_{gg}(\pTveto,\mu)
   &=& a_s \left[ \Gamma_0^A L_\perp + d_1^{\rm veto}(R) \right]
    + a_s^2 \left[ \Gamma_0^A\beta_0\,\frac{L_\perp^2}{2} + \Gamma_1^A\,L_\perp 
    + d_2^{\rm veto}(R) \right] \nonumber\\
   &&\mbox{}+ a_s^3 \left[ \Gamma_0^A\beta_0^2\,\frac{L_\perp^3}{3}
    + \left( \Gamma_0^A\beta_1 + 2\Gamma_1^A\beta_0 \right) \frac{L_\perp^2}{2} 
    + L_\perp \left(\Gamma _2^A+2 \beta _0\,d_2^{\rm veto}(R) \right) + d_3^{\rm veto}(R) \right] ,
   \nonumber\\
   h_A(\pTveto,\mu)
   &=& a_s \left[ \Gamma_0^A\,\frac{L_\perp^2}{4} - \gamma_0^g\,L_\perp \right]
    + a_s^2 \left[ \Gamma_0^A\beta_0\,\frac{L_\perp^3}{12}
    + \left( \Gamma_1^A - 2\gamma_0^g\beta_0 \right) \frac{L_\perp^2}{4}- \gamma_1^g\,L_\perp \right] ,
\end{eqnarray}
where we have defined the abbreviations $a_s=\alpha_s(\mu)/(4\pi)$ and $L_\perp=2\ln(\mu/\pTveto)$. The coefficients $\Gamma_n^A$, $\gamma_n^g$, and $\beta_n$ appear in the perturbative expansions of the anomalous dimensions and $\beta$-function, defined as
\begin{equation}
   \Gamma_{\rm cusp}^A(\mu) = \sum_{n=0}^\infty\,\Gamma_n^A\,a_s^{n+1} \,, \qquad
   \gamma^g(\mu) = \sum_{n=0}^\infty\,\gamma_n^g\,a_s^{n+1} \,, \qquad
   \beta(\mu) = -2\alpha_s(\mu) \sum_{n=0}^\infty\,\beta_n\,a_s^{n+1} \,.
\end{equation}

As long as the veto scale $\pTveto$ is in the perturbative domain, one can match the beam function $\bar B_g$ appearing in (\ref{colanom}) onto standard PDFs,
\begin{equation}\label{pdfmatch}
   \bar B_g(\xi,\pTveto)
   = \sum_{i=g,q,\bar q} \int_\xi^1\!\frac{dz}{z}\,
    \bar I_{g\leftarrow i}(z,\pTveto,\mu)\,\phi_{i/P}(\xi/z,\mu) \,,
\end{equation}
which is accurate up to hadronic corrections suppressed by powers of $\Lambda_{\rm QCD}/\pTveto$. The matching coefficients are connected by the simple rescaling relation
\begin{equation}
   \bar I_{g\leftarrow i}(z,\pTveto,\mu) 
   = e^{-h_A(\pTveto,\mu)}\,I_{g\leftarrow i}(z,\pTveto,\mu)
\end{equation}
to the functions $I_{g\leftarrow i}(z,\pTveto,\mu)$ computed at one-loop order in \cite{Becher:2012qa}. We find
\begin{equation}\label{Ires}
   \bar I_{g\leftarrow i}(z,\pTveto,\mu) 
   = \delta(1-z)\,\delta_{gi} + a_s \left[ - {\cal P}_{g\leftarrow i}^{(1)}(z)\,\frac{L_\perp}{2}
    + {\cal R}_{g\leftarrow i}(z) \right] + {\cal O}(a_s^2) , 
\end{equation}
where ${\cal P}_{g\leftarrow i}^{(1)}(z)$ are the one-loop DGLAP splitting functions. 

The explicit one-loop calculations of $F_{gg}$ and $I_{g\leftarrow i}$ performed in \cite{Becher:2012qa} show that (in the $\overline{\rm MS}$ scheme)
\begin{equation}
   d_1^{\rm veto}(R) = 0 \,, \qquad
   {\cal R}_{g\leftarrow g}(z) = - C_A\,\frac{\pi^2}{6}\,\delta(1-z) \,, \qquad
   {\cal R}_{g\leftarrow q}(z) = 2C_F z \,.
\end{equation}
At two-loop order, the anomaly coefficient $d_2^{\rm veto}(R)$ can be extracted from results presented in \cite{Banfi:2012jm}. One finds that
\begin{equation}\label{notCasi}
   d_2^{\rm veto}(R) = \left( \frac{808}{27} - 28\zeta_3 \right) C_A^2
    - \frac{224}{27}\,C_A T_F n_f - 32 C_A\,f(R) \,,
\end{equation}
where the expansion of $f(R)$ for small $R$ reads, in numerical form,\footnote{Except for the constant term, analytic expressions for the coefficients up to ${\cal O}(R^6)$ can be found in \cite{Banfi:2012yh}.} 
\begin{equation}\label{fR}
\begin{aligned}
   f(R) &= - \left( 1.0963\,C_A + 0.1768\,T_F n_f \right) \ln R
    + \left(  0.6106\,C_A -  0.0310\,T_F n_f \right) \\
   &\hspace{4.7mm}\mbox{}- \left( 0.5585\,C_A - 0.0221\,T_F n_f \right) R^2
    + \left( 0.0399\,C_A - 0.0004\,T_F n_f \right) R^4 + \dots \,.
\end{aligned}
\end{equation} 
In the following section we will reproduce this expression based on a two-loop calculation in SCET, which relies on the structure of the factorization formula (\ref{dsigdy}). The fact that we will reproduce the above expression exactly provides a non-trivial test of our factorization theorem at two-loop order. The three-loop coefficient $d_3^{\rm veto}(R)$ in (\ref{Fhexpansions}) is presently still unknown and will be estimated in Section \ref{sec:2loop} below, where we will also extract the two-loop corrections to the beam functions $\bar B_g(\xi_1,\pTveto)$ in (\ref{pdfmatch}) in numerical form.

We can now rewrite the jet-veto cross section from (\ref{dsigdy}) in the final, factorized form
\begin{equation}\label{sigfinal}
   \frac{d\sigma(\pTveto)}{dy}
   = \sigma_0(\pTveto)\,\bar H(m_t,m_H,\pTveto)\,\bar B_g(\xi_1,\pTveto)\,\bar B_g(\xi_2,\pTveto) \,,
\end{equation}
where we have introduced the RG-invariant hard function
\begin{equation}\label{Hbardef}
   \bar H(m_t,m_H,\pTveto) 
   = \left( \frac{\alpha_s(\mu)}{\alpha_s(\pTveto)} \right)^2 C_t^2(m_t^2,\mu) 
    \left| C_S(-m_H^2,\mu) \right|^2 \left( \frac{m_H}{\pTveto} \right)^{-2F_{gg}(\pTveto,\mu)}\,
    e^{2h_A(\pTveto,\mu)} \,,
\end{equation}
which contains all dependence on the short-distance scales $m_t$ and $m_H$. The dependence on rapidity is carried only by the beam functions $\bar B_g(\xi_{1,2},\pTveto)$. Note that, due to the collinear anomaly, it is not possible to factorize the dependence on the jet-veto scale $\pTveto$ in the hard function $\bar H$. However, it {\em is\/} possible to resum all large logarithms in the ratio $m_H/\pTveto$ consistently, to all orders in perturbation theory. To this end, one chooses a low factorization scale $\mu\sim\pTveto$ in the factorization formula (\ref{sigfinal}). Then the kernel functions $\bar I_{g\leftarrow i}$ required to compute the beam function $\bar B_g$ can be calculated in fixed-order perturbation theory. Likewise, the fixed-order expressions for $F_{gg}$ and $h_A$ in (\ref{Fhexpansions}) are sufficient. On the other hand, the matching coefficients $C_t$ and $C_S$ need to be computed in RG-improved perturbation theory. They can be evolved from the high matching scales $\mu\sim m_t$ and $\mu^2\sim-m_H^2$, where the matching calculations are performed, down to lower scales $\mu\sim\pTveto$ using RG equations. We will require the resulting expressions at next-to-next-to-leading order (NNLO) in RG-improved perturbation theory, which is equivalent to N$^3$LL accuracy. The corresponding expressions can be found in eqs.~(20) and (22) of \cite{Ahrens:2008nc}, with further details given in the Appendix of \cite{Becher:2006mr}.

All objects in the factorization formula (\ref{sigfinal}) are defined in a RG-invariant way, i.e.\ they are formally independent of the factorization scale $\mu$. As is common practice, we can use the residual $\mu$ dependence arising when the expressions (\ref{pdfmatch}) and (\ref{Hbardef}) are evaluated at some fixed order in perturbation theory as an indicator of the remaining perturbative uncertainties. This can be done for each of these objects separately, not just for the total cross section. We also note that the expression for the hard function becomes particularly simple if one adopts the default scale choice $\mu=\pTveto$ on the right-hand side of (\ref{Hbardef}). In this case
\begin{equation}
\begin{aligned}
   \bar H(m_t,m_H,\pTveto) 
   &= C_t^2(m_t^2,\pTveto) \left| C_S(-m_H^2,\pTveto) \right|^2 
    \left( \frac{m_H}{\pTveto} \right)^{-2F_{gg}(\pTveto,\pTveto)} \,, \\
   F_{gg}(\pTveto,\pTveto) &= \sum_{n=2}^\infty\,d_n^{\rm veto}(R)
    \left( \frac{\alpha_s(\pTveto)}{4\pi} \right)^n .
\end{aligned}
\end{equation}

\section{Jet clustering, multipole expansion, and zero bins}
\label{sec:multipole}

\begin{figure}[t!]
\begin{center}
\psfrag{s}{\Large $s$}\psfrag{c}{\Large $c$}\psfrag{a}{\Large $\bar{c}$}\psfrag{h}{\Large $h$}
\psfrag{p}[b]{\large $p_+$}\psfrag{m}[l]{\large $p_-$}
\psfrag{Q}[r]{$m_H$}\psfrag{L}[r]{$\lambda\, m_H$}\psfrag{U}[r]{$\lambda^2 m_H$}
\psfrag{q}[]{$\phantom{\lambda\,}m_H$}\psfrag{u}[]{$\lambda^2 m_H$}\psfrag{l}[]{$\lambda\, m_H$} \psfrag{o}{}
\includegraphics[width=0.45\textwidth]{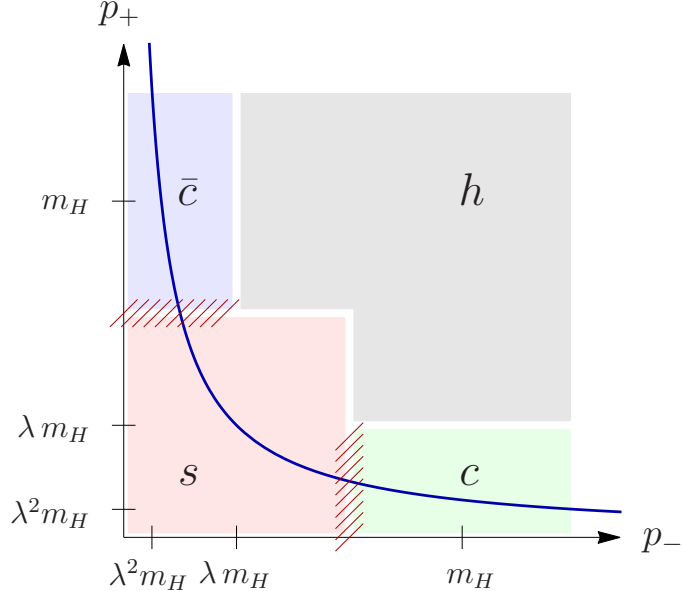}
\caption{\label{fig:regions} 
Momentum regions relevant for the jet-veto cross section. The figure indicates the scaling of the $p_+$ and $p_-$ components of soft ($s$), collinear ($c$), anti-collinear ($\bar c$), and hard ($h$) contributions. The hyperbola corresponds to $p_+p_-=(\pTveto)^2=\lambda^2 m_H^2$. The red hatching shows the soft-collinear overlap regions.}
\end{center}
\end{figure}

We now analyze the factorization properties of the jet-veto cross section using the formalism of SCET, in which highly energetic particles aligned with the colliding protons are described in terms of collinear and anti-collinear quark and gluon fields, and soft particles emitted from the beam jets are described in terms of soft fields. The effective theory implements an expansion of scattering amplitudes in powers of the small parameter $\lambda\sim\pTveto/m_H$, where the jet veto sets the characteristic size of all transverse momenta in the process. We introduce two light-like reference vectors $n^\mu$ and $\bar n^\mu$ (satisfying $n\cdot\bar n=2$) parallel to the beam axis and decompose all 4-vectors in the light-cone basis spanned by these vectors,
\begin{equation}\label{lightcone}
   p^\mu = n\cdot p\,\frac{\bar n^\mu}{2} + \bar n\cdot p\,\frac{n^\mu}{2} + p_\perp^\mu 
   \equiv p_+\,\frac{\bar n^\mu}{2} + p_-\,\frac{n^\mu}{2} + p_\perp^\mu \,.
\end{equation}
The different types of modes relevant to our discussion are characterized by the scalings of their momenta $(p_+,p_-,p_\perp)$ with powers of $\lambda$, namely $p_c^\mu\sim m_H(\lambda^2,1,\lambda)$ for collinear particles, $p_{\bar c}^\mu\sim m_H(1,\lambda^2,\lambda)$ for anti-collinear particles, and $p_s^\mu\sim m_H(\lambda,\lambda,\lambda)$ for soft particles. Hence, the particles in these three categories have transverse momenta of order the jet veto, but very different rapidities. The scaling of these modes is displayed graphically in Figure~\ref{fig:regions}. In addition, the cross section receives contributions from the hard momentum region $p_h^\mu\sim m_H(1,1,1)$, where we do not distinguish between $m_H$ and $m_t$. These corrections are purely virtual and are integrated out in the construction of the effective theory. One may also worry about the contributions from modes with smaller virtualities, $p^2\ll(\pTveto)^2$. For example, an on-shell soft mode, which accidentally is closely aligned with the beam axis, would have momentum scaling $\sim m_H(\lambda^2,\lambda,\lambda^{3/2})$. This mode has a rapidity lying in between that of collinear and soft modes. Indeed, it may also be regarded as a collinear mode whose minus component is accidentally small. The important point is that, because of their small transverse momenta, such modes play no role for the total transverse momentum of a jet. Therefore, an arbitrary number of them can be emitted, and their effect cancels out in the factorization theorem. This is in analogy with the cancellation of ultrasoft modes in the factorization theorem for the Drell-Yan cross section at small transverse momentum~\cite{Becher:2010tm}.

As explained in \cite{Becher:2012qa}, the jet clustering algorithm does not group particles with different momentum scalings (collinear, anti-collinear, or soft) into the same jet. The reason is that, generically, the rapidity difference between two such particles are such that $\Delta y_{ij}\sim\ln(m_H/\pTveto)$, which by assumption is much larger than $R$, see (\ref{R1}). As a consequence, in the jet algorithm (\ref{algo}) the distance measure $d_{ij}$ for two such particles is always larger than the minimum of $d_{iB}$ and $d_{jB}$. Since the soft and (anti-)collinear modes have the same virtuality, they live along the hyperbola in the $(p_+,p_-)$ plane shown in Figure \ref{fig:regions}, and their precise separation along this hyperbola is to some extent arbitrary. The fact that these modes differ by large rapidities gives rise to large logarithms, which are accounted for by the collinear anomaly. In complete analogy with the construction of the SCET Lagrangian, where based on the {\em generic\/} scalings of the fields one does not include soft-collinear interaction terms, it is unnecessary to consider the degenerate case where a collinear and a soft mode near the boundary are clustered into a single jet. Since there are no enhancements of the cross section in these power-suppressed phase-space regions, boundary effects do not contribute at leading power. Only in corners of the phase space, e.g.\ when a soft emission becomes collinear to the beam, soft and collinear radiation can be clustered into the same jet. However, since the cross section does not exhibit additional singularities in the corresponding limit, such configurations only give rise to power-suppressed contributions. 

The argument just presented has been challenged in \cite{Tackmann:2012bt}, where it was argued that the clustering of soft and collinear modes near the boundary of phase space that separates them in rapidity does give rise to leading-order contributions to the cross section starting at NNLL order, unless the jet-radius parameter $R$ is taken to be parametrically much smaller than~1. If this was true, then there would be no region in parameter space where our factorization theorem (\ref{sigfinal}) would be useful, since for parametrically small values of $R$ it does not accomplish the resummation of $\ln^n\!R$ terms. The argument presented in \cite{Tackmann:2012bt} was backed up by a calculation of a particular soft-collinear clustering contribution, which was found to be non-zero and provided a leading-power contribution to the cross section proportional to $R^2$ and $R^4$, hence the claim that these contributions are suppressed only for parametrically small $R$. We will now demonstrate in detail that these findings are not in conflict with our factorization formula.

SCET is based on the method of regions. Loop and phase-space integrations are split into different momentum regions in a systematic manner dictated by the structure of the effective Lagrangian. One could separate the different regions using cutoffs, as indicated graphically in Figure~\ref{fig:regions}, but this is impractical because it would spoil gauge invariance in the individual sectors of the effective theory. Instead, one uses dimensional regularization to handle the appearing singularities. It is then crucially important to perform the calculation of SCET diagrams using a multipole expansion, which expands out components of particle momenta (or position vectors) that are parametrically suppressed with respect to the leading ones in a given interaction or propagator \cite{Beneke:2002ph}. Only if this is done consistently, there is no double counting of  momentum  configurations. The reason is that, once we consider the contribution from a certain momentum region to an integral, any expansion of the integrand around another limit will leave us with a scaleless integral, which is zero in dimensional (or, more generally, analytic) regularization. In the present case, it is important that one performs the multipole expansion not only for the integrands of loop or phase-space integrals, but also for the measurement functions ${\cal M}_{\rm veto}$ in the definitions (\ref{collfun}) and (\ref{softfun}).

If one does not perform the multipole expansion consistently, then there arise contributions from the double counting of overlapping momentum regions, which must be subtracted by hand in order to obtain the correct result. In the SCET community these subtraction terms are referred to as ``zero-bin subtractions'' \cite{Manohar:2006nz}. The problem with the argument presented in \cite{Tackmann:2012bt} is that the soft-collinear clustering contribution was calculated {\em without\/} performing the multipole expansion, but the relevant zero-bin subtractions were not evaluated. We will argue that these zero-bin subtractions exactly cancel the soft-collinear clustering term, so that one recovers the same result as before. When the multipole expansion is performed consistently, the contributions in which different modes are clustered by the jet algorithm are simply zero.

We will now illustrate these statements with the help of a simple example, which demonstrates our point without involving the technical complexities of the real calculation. In Section~\ref{sec:2loop}, we will perform the calculation of the two-loop anomaly coefficient $d_2^{\rm veto}(R)$ in the context of SCET, considering only the contributions of two collinear or two soft emissions, in accordance with our factorization formula (\ref{dsigdy}), according to which the jet veto must be applied separately in each sector of SCET. The fact that in this way we reproduce the result extracted from \cite{Banfi:2012jm} proves that there are no missing contributions at this order. We thus do not confirm the statement made in \cite{Tackmann:2012bt} that soft-collinear mixing terms contribute to the cross section at ${\cal O}(\alpha_s^2)$. Moreover, since our arguments are completely general and not tied to a particular order in the perturbative expansion, they support our claim that the factorization formula (\ref{sigfinal}) remains valid also in higher orders in perturbation theory.

Consider the following simple rapidity integral with integrand 1 and the constraint that two particles with rapidities $y_c$ and $y$ are clustered into one jet:
\begin{equation}\label{simpex}
   I = \int_{-\infty}^{\infty}\!dy\,\theta\big(R^2-(y-y_c)^2\big) = 2R \,.
\end{equation}
Here $y_c\gg 1$ is the fixed rapidity of a collinear particle, and we assume that $R={\cal O}(1)$. The $\theta$-function plays the role of the measurement function ${\cal M}_{\rm veto}$ in the definitions (\ref{collfun}) and (\ref{softfun}). Even though this integral is extremely simple, it captures the main features of the integrals we will encounter in the computation of the $C_F^2$ term in Section~\ref{sec:CF2}, since the relevant part of the amplitude for this color structure only depends on the transverse momentum. The only difference to the trivial example integral (\ref{simpex}) is that one also integrates over the rapidity of the second emission and the azimuthal angles, which then also enter the $\theta$-function constraint. In the context of SCET we should evaluate the integral as a sum over contributions from different momentum regions. In each region we must multipole expand the argument of the $\theta$-function according to the rules of the effective theory. For the purposes of our discussion we will consider for the moment only the contributions from collinear and soft partons; as we discuss below, adding the anti-collinear region would not change the argument. A collinear particle has momentum scaling $(\lambda^2,1,\lambda)$. In the collinear region both rapidities scale the same, $y\sim y_c\sim\ln(1/\lambda)\gg 1$, but their difference is ${\cal O}(1)$. There is thus nothing to expand in the argument of the $\theta$-function, and we get $I_c=I$ for the collinear-collinear clustering term. A soft particle has momentum scaling $(\lambda,\lambda,\lambda)$ and hence $y={\cal O}(1)$. It follows that in the argument of the $\theta$-function $(y-y_c)^2={\cal O}(\ln(1/\lambda))$ is parametrically larger than $R^2={\cal O}(1)$. We must therefore perform the multipole expansion
\begin{equation}\label{thetaexp}
   \theta\big(R^2-(y-y_c)^2\big) 
   = \theta\big(-(y-y_c)^2\big) + R^2\,\delta\big((y-y_c)^2\big) + \dots \,.
\end{equation}
The higher-order terms in the expansion will contain derivatives of $\delta$-functions, but because the arguments are always non-zero the entire expression on the right-hand side just vanishes.  

At this point one may worry about double counting, since in the collinear-collinear contribution we have also integrated over the region where the collinear particle becomes soft. One should therefore subtract the contribution from the soft-collinear overlap region (the ``zero bin''), which otherwise would be counted twice. However, after the multipole expansion this overlap contribution vanishes, $I_{(cs)}=0$, for the same reason that the soft-collinear contribution vanishes. Alternatively, one could evaluate the contributions from the two regions without performing the multipole expansion. Then obviously both regions yield the same contribution, $I_c=I_s=I$. But now the double-counted soft-collinear overlap contribution is also non-zero, and indeed $I_{(cs)}=I$ is equal to the soft-collinear contribution. The final result is $I_c+I_s-I_{(cs)}=I$, as it should be. In analogy with the findings of \cite{Tackmann:2012bt} the soft-collinear clustering term is non-zero in this case, but its contribution is precisely cancelled by the zero-bin subtraction. 

When also the anti-collinear region is included, we still only need to compute the contribution $I_c$ if the multipole expansion is performed consistently, but when working with subtractions the procedure gets more complicated. The general expression for three momentum regions reads
\begin{equation}\label{subtr}
   I = I_c + I_s + I_{\bar c} - I_{(cs)} - I_{(\bar c s)} - I_{(\bar c c)} + I_{(\bar c cs)} \,.
\end{equation}
The last term $I_{\bar c cs}$ describes the double overlap region, where the momentum can simultaneously be part of any region. It is obtained by expanding the integrand in the limit where the momentum scales as $(\lambda^2,\lambda^2,\lambda)$. It has to be added back, since the other three subtractions would remove this region from the integral. The general systematics of subtractions was studied in detail in \cite{Jantzen:2011nz}, as a step towards a proof of the method of regions. In our simple example, all of the above contributions are equal to the original integral $I$. Since the momenta in the double overlap region and in the $(\bar c c)$ contribution scale in the same way, the last two terms in (\ref{subtr}) are identical for any given integral, and the general expression simplifies to 
\begin{equation}\label{subtrsimp}
   I = I_c + I_s +  I_{\bar c}  - I_{(cs)} - I_{(\bar c s)} \,.
\end{equation}
While useful to map the integrals in dimensional regularization onto standard integrals, the subtraction procedure is extremely cumbersome in practice. For the two-emission case, for example, one would start off with 25 momentum configurations, since each of the two momenta can be in any of the regions or overlap regions in (\ref{subtrsimp}). In addition to the proliferation of regions, another drawback of the subtraction method is fact that the integrals are no longer homogenous in the expansion parameter $\lambda$, so that in general one will need to reexpand the final result in $\lambda$ after integrating.

It may appear strange at first sight that we had to expand the argument of the $\theta$-function in (\ref{simpex}) in powers of $\ln(1/\lambda)$, not in powers of $\lambda$. This distinction is however meaningless. Instead of (\ref{thetaexp}) we may equally well write $\theta(e^{-|y-y_c|}-e^{-R})=\theta(-e^{-R})+{\cal O}(\lambda)$, where $e^{-|y-y_c|}={\cal O}(\lambda)$ for a soft particle. The multipole expansion is now an expansion in powers of $\lambda$. Indeed, one can always rewrite the rapidity integrals in terms of integrals over components of light-cone momenta. For example, denoting the collinear reference momentum by $k$ and the soft momentum by $p$, we have $y_c=\ln(k_+/k_T)$ and $y=\ln(p_+/p_T)$, and hence the phase-space constraint can be rewritten in the form
\begin{equation}\label{eq:phasespace}
\begin{aligned}
   \theta\big(R^2-(y-y_c)^2\big) 
   &= \theta\big(R-(y-y_c)\big)\,\theta(y-y_c) + \theta\big(R-(y_c-y)\big)\,\theta(y_c-y) \\
   &\hspace{-3.3cm}
    = \theta(\underbrace{e^R p_T k_+}_{\lambda^3}-\underbrace{p_+ k_T}_{\lambda^2})\,
    \theta(\underbrace{p_+ k_T}_{\lambda^2}-\underbrace{p_T k_+}_{\lambda^3})
    + \theta( \underbrace{p_+ k_T}_{\lambda^2}-\underbrace{e^{-R} p_T k_+}_{\lambda^3})\,
    \theta(\underbrace{p_T k_+}_{\lambda^3}-\underbrace{p_+ k_T}_{\lambda^2}) \,.
\end{aligned}
\end{equation}
In the last step we have indicated the scalings of the various soft and collinear momentum components. Neglecting higher-order terms in $\lambda$, we obtain
\begin{equation}
   \theta\big(R^2-(y-y_c)^2\big) 
   = \theta(-p_+ k_T)\,\theta(p_+ k_T) + \theta(p_+ k_T)\,\theta(-p_+ k_T) + \dots \,,
\end{equation}     
which vanishes, since each light-cone component of an on-shell momentum is positive. We can further think of the $\theta$-functions of momentum components as the discontinuities of some propagators. This clearly shows that the multipole expansion in (\ref{thetaexp}) is not different from multipole expansions of propagators in ordinary SCET loop or phase-space integrals, and it makes it clear that power-suppressed terms, which are expanded out, are governed by powers of $\lambda$, not powers of $1/\ln(1/\lambda)$. 

We finish this section with an important remark. The structure of the first $\theta$-function in (\ref{eq:phasespace}) suggests that some of the power-suppressed terms may be accompanied by a factor $e^R$. Because we treat $R$ as an ${\cal O}(1)$ parameter, also $e^R$ is {\em not\/} a parametrically large quantity, so even if such terms exist, their presence would not upset the structure of the factorization formula (\ref{sigfinal}). The question of the numerical size of power-suppressed corrections must be separated from the issue of parametrically enhanced corrections. The outcome of our discussion is that, for $R={\cal O}(1)$, there are {\em no contributions\/} to the cross section arising from soft-collinear clustering terms, which would upset the factorization formula. In our framework all soft contributions are purely scaleless. In physical terms, this means that the soft contributions can effectively be absorbed into the (anti-)collinear fields. The structure of relation (\ref{colanom}) implies that this is indeed possible. The same happens for the transverse-momentum spectrum of electroweak bosons \cite{Becher:2012qa,Becher:2010tm}, and also in all SCET$_{\rm I}$ applications where a separate mode with $(\lambda,\lambda,\lambda)$ scaling is not needed to describe the physics. Nevertheless, there are power-corrections to our factorization formula from subleading terms in the effective Lagrangian and subleading SCET operators. In Section~\ref{sec:numerics} we will study their numerical impact by matching our results to the cross section computed in fixed-order perturbation theory. We will find that even for $R=1$ the power corrections remain small; indeed, we will not find any numerical evidence for the existence of $e^R$-enhanced power corrections.

\section{Two-loop computation of the anomaly exponent}
\label{sec:2loop}

\begin{figure}
\begin{center}
\includegraphics[width=1.0\textwidth]{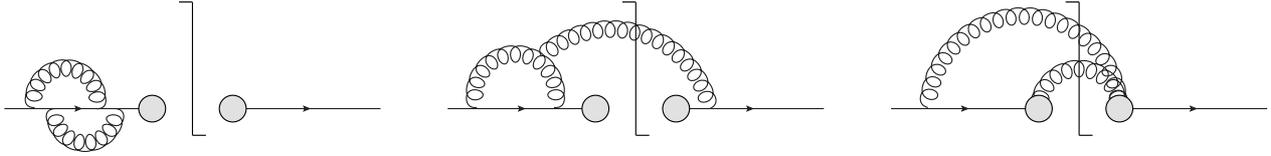}
\caption{\label{fig:collcollrealvirt} 
Examples of two-loop diagrams contributing to the collinear function ${\cal B}_{c,q}(z,\pTveto,\mu)$. The collinear Wilson lines are denoted by the gray blobs.}
\end{center}
\end{figure}

We now turn to the computation of the two-loop anomaly exponent $d_2^{\rm veto}(R)$ in (\ref{Fhexpansions}). According to the factorization formula (\ref{sigfinal}), this quantity can be obtained from a perturbative computation of the collinear and soft matrix elements defined in (\ref{collfun}) and (\ref{softfun}). Instead of the beam function ${\cal B}_{c,g}$ for a gluon, we will in the following consider the analogous function for a collinear quark, defined as
\begin{equation}\label{collfunquark}
   {\cal B}_{c,q}(z,\pTveto,\mu)
   = \int\frac{dt}{(2\pi)}\,e^{-izt\bar n\cdot p}\,      
    \sum \hspace{-0.8cm}\int\limits_{X_c,\,{\rm reg.}}\!\! 
    {\cal M}_{\rm veto}(\pTveto,R, \{ \underline{p_c} \})\,
    \langle P(p)|\,\bar\chi_c(t\bar n)\,|X_c\rangle\,\langle X_c|\,\chi_c(0)\,|P(p)\rangle \,.
\end{equation}
This function would appear in the calculation of the jet-veto cross section for a quark-initiated process such as $Z$-boson production at the LHC. As we will explain below, the result for $d_2^{\rm veto}(R)$ relevant for Higgs production can be obtained from the corresponding coefficient for $Z$ production by replacing $C_F\to C_A$, but distinguishing the two color factors will make it easier to organize the calculation. The gauge-invariant gluon and quark fields in the matrix elements in (\ref{collfun}) and (\ref{collfunquark}) are related to the usual QCD fields by \cite{Bauer:2001yt,Bauer:2002nz,Hill:2002vw}
\begin{equation}\label{eq:collfields}
   g_s\,{\cal A}_{c\perp}^\mu(x) = W^\dagger(x)\,[iD_\perp^\mu W(x)] \,, \qquad
   \chi_c(x) =\frac{n\!\!\!/\bar n\!\!\!/}{4}\,W^\dagger(x)\,\psi(x) \,,
\end{equation}
where $W(x)$ is a straight Wilson line along the $\bar n$ direction from $-\infty$ to $x$.  We use the standard QCD Lagrangian to evaluate the collinear matrix element in (\ref{collfunquark}), since the collinear SCET Lagrangian is equivalent to it (see e.g.\ \cite{Beneke:2002ph,Becher:2006qw}). Some representative examples of two-loop diagrams contributing to this matrix element are shown in Figure~\ref{fig:collcollrealvirt}. 

In order to extract the anomaly coefficient $d_2^{\rm veto}(R)$, we can evaluate the collinear matrix elements (\ref{collfun}) and (\ref{collfunquark}) with partonic instead of hadronic external states. In addition, we also need to calculate the soft function defined in (\ref{softfun}), which involves products of soft Wilson lines along the two beam directions. For the Higgs case, these are Wilson lines in the adjoint representation, while the fundamental representation is relevant for the case of $Z$-boson production. The normalization factor becomes $d_R=N_c$ in the latter case, such that ${\cal S}(\pTveto,\mu)=1$ at lowest order in perturbation theory. In the analytic regularization scheme based on the prescription (\ref{analytreg}), the soft function is given by scaleless integrals of the type
\begin{equation}
   \int_{-\infty}^{\infty}\!dy_t\,e^{-\alpha y_t}\equiv 0 \,,
\end{equation}
which vanish by definition. The reason is that the integral over the total rapidity $y_t$ of the emitted soft gluons is not constrained by the jet veto. It follows that ${\cal S}(\pTveto,\mu)=1$ to all orders in this regularization scheme. In principle, it is thus sufficient to evaluate the collinear functions ${\cal B}_{c,q}$ and ${\cal B}_{c,\bar q}$, and since the divergences in the analytic regulator must cancel in the product of these functions, calculating the left- or right-collinear function would be  sufficient in practice.

In our calculation we will, however, adopt a different strategy. It has been shown in \cite{Becher:2012qa} that one obtains a non-zero soft function if one imposes different jet vetoes for the left- and right-moving particles. The anomalous large logarithms in the soft function are then tied, via the anomaly equations, to those in the collinear beam functions. Extracting the coefficient $d_2^{\rm veto}(R)$ from the soft function offers the advantage that the relevant Wilson-line diagrams are simpler to compute than the loop diagrams for the collinear functions. Instead of imposing different jet vetoes for left- and right-moving particles, we can generate a non-trivial soft function by using different analytic regulators for them. To this end, we generalize the regularization prescription in (\ref{analytreg}) by replacing
\begin{equation}\label{splitreg}
   \left( \frac{\nu}{k_+} \right)^\alpha 
   \to \left( \frac{\nu}{k_+} \right)^\alpha \theta(k_+ - k_-)
    + \left( \frac{\nu}{k_+} \right)^\beta \theta(k_- - k_+) \,.
 \end{equation}
After the multipole expansion, the collinear function only involves the regulator $\beta$, while the anti-collinear function is regularized by $\alpha$. The cancellation of divergences between the soft and (anti-)collinear functions then proceeds in the way shown schematically in Table~\ref{table:corrmat}. Because of the structure of the cancellations, the computation of the divergence of a single function is again sufficient, and with the regulator (\ref{splitreg}) we can work with the soft instead of the (anti-)collinear functions. For convenience, we will perform the extraction of the color structures $C_F C_A$ and $C_F T_F n_f$ of the anomaly coefficient $d_2^{\rm veto}(R)$ from the computation of the soft function using the split regulator (\ref{splitreg}), while we will extract the $C_F^2$ part from the collinear functions with the original form (\ref{analytreg}) of the regulator. 

\begin{table}[t!]
\centering
\renewcommand{\arraystretch}{1.8}
\begin{tabular}{ccccc}
$cc$ && $ss$ && $\bar c\bar c$ \\ 
\midrule
$\frac{1}{\beta}\left(\frac{\nu\,m_H}{(p_T^{\mathrm{veto}})^2}\right)^{2\beta}$
 && $-\frac{1}{\beta}\left(\frac{\nu}{p_T^{\mathrm{veto}}}\right)^{2\beta}
    + \frac{1}{\alpha}\left(\frac{\nu}{p_T^{\mathrm{veto}}}\right)^{2\alpha}$
 && $-\frac{1}{\alpha}\left(\frac{\nu}{m_H}\right)^{2\alpha}$ \\[6pt]
\end{tabular}
\caption{\label{table:corrmat}
Structure of the divergences of two-emission diagrams arising when one uses the regulator (\ref{splitreg}), which distinguishes left- and right-moving particles. We denote collinear particles by $c$, anti-collinear ones by $\bar{c}$, and soft ones by $s$.}
\end{table}

A second simplification of the computation is achieved by using the fact that the two-loop anomaly coefficient for the transverse-momentum spectrum of an electroweak boson $B$ (with $B=H,Z,\gamma^*,W^\pm$) is known \cite{Becher:2010tm}. Since the jet algorithm only has an effect for two and more emissions, the difference 
\begin{equation}\label{sigdif}
   \Delta\hat\sigma(\pTveto) 
   = \frac{1}{\sigma_0}\,\big[ \sigma_{\rm veto}(\pTveto) - \sigma_B(\pTveto) \big]
\end{equation}
between the jet-veto cross section $\sigma_{\rm veto}(\pTveto)$ and $\sigma_B(\pTveto)$, the boson $q_T$ spectrum integrated up to a momentum scale $\pTveto$, starts at ${\cal O}(\alpha_s^2)$ and involves only contributions from two real-emission diagrams at this order. This observation was used in \cite{Banfi:2012yh} to extract the $R$-dependent part of the NNLL order corrections, and the logarithmically-enhanced terms in the above difference were given explicitly in \cite{Banfi:2012jm}. On the partonic level, the logarithmically-enhanced two-loop terms have the form
\begin{equation}\label{Deltad2}
   \Delta\hat\sigma(p_T) = -2\delta(\hat s-m_H^2)\,\left(\frac{\alpha_s}{4\pi}\right)^2 
   \Delta d^{\rm veto}_2(R)\,\ln\frac{m_H}{\pTveto} + \dots \,,
\end{equation}
where $\hat{s}$ is the partonic center-of-mass energy squared. In order to obtain the full two-loop anomaly coefficient, we then use the relation \cite{Becher:2012qa}
\begin{equation}\label{bosontoveto}
   d_2^{\rm veto}(R) = d_2^B + 32\zeta_3 C_B^2 + \Delta d^{\rm veto}_2(R) \,,
\end{equation}
where $C_B=C_A$ for Higgs production and $C_B=C_F$ for $Z$-boson production. The $\zeta_3$ term arises from the Fourier integral present in the factorization formula for the boson $q_T$ spectrum. The anomaly coefficient relevant for the transverse-momentum spectrum,
\begin{equation}
   d_2^B = C_B \left[ \left(\frac{808}{27}  - 28\zeta_3 \right) C_A
    - \frac{224}{27}\,T_F n_f\right] ,
\end{equation}
was extracted in \cite{Becher:2010tm}. We will find that the quantity $\Delta d^{\rm veto}_2(R)$ defined in (\ref{Deltad2}) can be further decomposed as
\begin{equation}\label{Deltadd2}
   \Delta d_2^{\rm veto}(R) = - 32\zeta_3 C_B^2 - 32 C_B\,f(R) \,,
\end{equation}
where $f(R)$ vanishes for $R\to\infty$, and for $R<\pi$ it can be approximated by the numerical expression given in (\ref{fR}). In this way, we recover the result (\ref{notCasi}) once we set $C_B=C_A$ for the Higgs-boson case.

The real-emission QCD diagrams contributing to $\Delta\hat\sigma(\pTveto) $ are free of infrared singularities and can be evaluated in $d=4$ space-time dimensions. However, the effective-theory diagrams will continue to suffer from light-cone divergences, and thus the analytic regularization has to be kept. The two-emission measurement function relevant for the difference $\Delta\hat\sigma(\pTveto)$ reads
\begin{eqnarray}
   {\cal M}_\Delta(\pTveto)
   &=& {\cal M}_{\rm veto}(\pTveto) - {\cal M}_B(\pTveto) \nonumber\\
   &=& \theta(R-\Delta R)\,\theta(\pTveto-p_T)
    + \theta(\Delta R-R)\,\theta(\pTveto-k_T)\,\theta(\pTveto-q_T) - \theta(p_T^{\rm veto}-p_T) 
    \nonumber\\
   &=& \theta(\Delta R-R)\,\Big[ \theta(\pTveto-k_T)\,\theta(\pTveto-q_T)
    - \theta(\pTveto-p_T) \Big] \,,
\end{eqnarray}
where $p=k+q$ is the total momentum of the two emissions with momenta $k$ and $q$, and $\Delta R=\sqrt{\Delta y^2+\Delta\phi^2}$ is their angular separation. 

There is a price one has to pay when working with the difference $\Delta\hat\sigma(\pTveto)$ instead of the jet-veto cross section itself. With the measurement function $\mathcal{M}_{\Delta}(\pTveto)$, contributions arise from clustered particles that have large angular separations $\Delta R > R$. In contrast to the jet-veto cross section, $\Delta\hat\sigma(\pTveto)$ {\em does\/} get contributions from particles from the different sectors and we will therefore need to evaluate those contributions. The physics reason is that collinear and soft particles both contribute equally to the $q_T$ spectrum of the electroweak boson. The mixing contributions are $R$ independent and only arise for the $C_F^2$ color structure. Their presence is the reason why we evaluate this part with the  the standard form (\ref{analytreg}) of the analytic regulator, for which the soft region is absent. We then only need to compute the mixing contribution involving one collinear and one anti-collinear particle.

\subsection{\boldmath Evaluation of the $C_F C_A$ and $C_F T_F n_f$ terms}

To extract the contribution of these two color structures to $d_2^{\rm veto}(R)$, we compute the two-loop soft function with the split analytic regulator (\ref{splitreg}). Up to the choice of the regulator, the corresponding computation is identical to what was done in \cite{Tackmann:2012bt}, with one important difference: this paper claimed that the factorization of the cross section would only hold for $R\to 0$ and the computation was only performed in this limit. As we have demonstrated in Section~\ref{sec:multipole}, the factorization formula (\ref{sigfinal}) also holds at finite $R={\cal O}(1)$, and we must therefore recover the full QCD result of \cite{Banfi:2012yh} from our computation of the soft function. 

In order to perform the calculation, one needs the two-emission soft amplitude squared,
\begin{equation}
   {\cal A}_s(k,l) = \sum_{\rm pol.}\, \left| {\cal M}_{2g}(k,l) \right|^2 ,
\end{equation}
which is given in compact form in Appendix~C of \cite{Becher:2012qc} and can also be found in \cite{Tackmann:2012bt}. One then parameterizes the integration over the two-particle phase space in terms of angles, rapidities, and transverse momenta, introducing the variables
\begin{equation}\label{eq:kin}
   \Delta y = y_k - y_l \,, \qquad 
   \Delta\phi = \phi_k - \phi_l \,, \qquad
   p_T =k_T +l_T \,, \qquad
   z = \frac{k_T}{p_T} \,.
\end{equation} 
The integration over the total rapidity $y_t$ then gives rise to a divergence of the form
\begin{equation}
   - \frac{1}{\beta} \left( \frac{\nu}{\pTveto} \right)^{2\beta}
   + \frac{1}{\alpha} \left( \frac{\nu}{\pTveto} \right)^{2\alpha} ,
\end{equation}
whose coefficient is the collinear anomaly. The divergence only arises if both emissions are either to the left or two the right, and the two terms would cancel if we were to set $\alpha=\beta$. The integration over $p_T$ can be performed analytically, which leads to the result
\begin{equation}\label{softcorr}
\begin{aligned}
   \Delta\hat\sigma(\pTveto) 
   &=\delta(\hat s-m_H^2) \left[ \frac{2}{\alpha} \left( \frac{\nu}{\pTveto} \right)^{2\alpha}
    - \frac{2}{\beta} \left( \frac{\nu}{\pTveto} \right)^{2\beta} \right] \\
   &\quad\times \int_0^1\!dz\,\int_{-\infty}^{\infty}\!d\Delta y\,\int_0^\pi\frac{d\Delta\phi}{\pi}\,
    \frac{1}{(4\pi)^4}\,\theta\big(\sqrt{\Delta y^2+\Delta\phi^2}-R\big) \\
   &\quad\times \left[ (\pTveto)^4\,z(1-z)\,{\cal A}_s(k, l) \right] 
    \ln\frac{\sqrt{z^2+(1-z)^2+2z(1-z)\cos\Delta\phi}}{\mbox{max}(z,1-z)} \,.
\end{aligned}
\end{equation}
For a given value of $R$, the remaining integrations can be performed numerically. To obtain an analytic form of the result, we have expanded the integrand in powers of $R$, as was done in \cite{Banfi:2012yh}. Details of the calculation can be found in Appendix~\ref{App:correlated}. Translating the divergence in the analytic regulator into the anomalous logarithm according to the structures shown in Table~\ref{table:corrmat}, and using relation (\ref{Deltad2}), we obtain \begin{equation}
\begin{aligned}
   \Delta d_2^{\rm veto}(R) \big|_{C_F C_A,\,C_F T_F n_f}
   &= - 32 C_F C_A\,\Big( c_L^A \ln R + c_0^A + c_2^A R^2 + c_4^A R^4 + \dots \Big) \\
   &\quad\mbox{}- 32 C_F T_F n_f\,\Big( c_L^f \ln R + c_0^f + c_2^f R^2 + c_4^f R^4 + \dots \Big) \,,
\end{aligned}
\end{equation}
where the first few expansion coefficients are given by
\begin{equation}\label{cLtoc4}
\begin{aligned}
   c_L^A &= \frac{131}{72} - \frac{\pi^2}{6} - \frac{11}{6} \ln 2 \,, &
    c_L^f &= - \frac{23}{36} + \frac{2}{3} \ln 2 \,, \\
   c_0^A &= -\frac{805}{216} + \frac{11 \pi^2}{72} + \frac{35 }{18}\ln 2 + \frac{11}{6} \ln^2 2+\frac{\zeta_3}{2} \,, \quad &
    c_0^f &= \frac{157}{108} - \frac{\pi^2}{18} -\frac{8}{9}\ln 2 - \frac{2}{3} \ln^22 \,, \\
   c_2^A &= \frac{1429}{172800} + \frac{\pi^2}{48} + \frac{13}{180} \ln 2 \,, &
    c_2^f &= \frac{3071}{86400} - \frac{7}{360} \ln 2 \,. 
\end{aligned}
\end{equation}
In Appendix~\ref{App:correlated}, we present analytic expressions for the expansion coefficients up to ${\cal O}(R^{10})$. Our results for the coefficients $c_L^i$ and $c_n^i$ with $n=2,4,6$ agree with the findings of \cite{Banfi:2012yh}. Our analytic expressions for the coefficients $c_0^i$ are new, and they are in agreement with the numerical values reported in \cite{Banfi:2012yh}.\footnote{There is a slight deviation for the $C_A$ part of the constant term, where we have $-c_0^A/c_L^A=\ln(1.7455)$, while \cite{Banfi:2012yh} quotes $\ln(1.74)$.}  

\subsection{\boldmath Evaluation of the $C_F^2$ term}
\label{sec:CF2}

The computation of the $C_F^2$ part is complicated by the fact that quadratic divergences in the analytic regulator as well as mixing terms between the different sectors arise. Rewriting $\theta(\Delta R-R)=1-\theta(R-\Delta R)$, we note that both problems do not affect the second term on the right, which can thus be treated exactly in the same way as the $C_F C_A$ and $C_F T_F n_f$ contributions studied in the previous section. This second term, which contains all $R$ dependence, corresponds to the independent-emission piece computed in \cite{Banfi:2012yh}. Proceeding in the same way as above, we confirm their result 
\begin{equation}
   \Delta d_2^{\rm veto}(R) \big|_{C_F^2}^{R\rm{-dep.}}
   = C_F^2 \left( \frac{8\pi^2 R^2}{3} - 2 R^4 \right) .
\end{equation}

This leaves the $R$-independent piece arising from the rewriting of the $\theta$-function. As discussed earlier, this part involves the mixing between the different sectors of the effective theory, but only because we consider the cross-section difference in (\ref{sigdif}) instead of the jet-veto cross section itself. We will compute it using the standard form (\ref{analytreg}) of the analytic regulator, for which the soft contributions are absent. The general form of this contribution in terms of rapidity, azimuthal angle, and transverse momentum is
\begin{equation}\label{eq44}
\begin{aligned}
   \Delta\hat\sigma_{ij}(\pTveto) \big|_{R\rm{-indep.}}
   &= \frac{1}{(16\pi^2)^2}\int_0^\infty\!dk_+ \int_0^\infty\!dk_- \int d^2k_\perp\,\delta(k^2) 
    \int_0^\infty\!dl_+ \int_0^\infty\!dl_- \int d^2l_\perp\,\delta(l^2) \\
   &\quad\times \left(\frac{\nu^2}{k_+ l_+} \right)^\alpha {\cal A}_{ij}(k,l)\, 
    \Delta_{ij}(k,l)\,{\cal M}_{R\rm{-indep.}}(k,l,\pTveto) \,, \quad
    \mbox{with $i,j=c,\bar c$.}
\end{aligned}
\end{equation}
Here ${\cal A}_{ij}(k,l)$ is the squared amplitude for two emissions in the appropriate momentum regions. The measurement function
\begin{equation}
   {\cal M}_{R\rm{-indep.}}(k,l,\pTveto) 
   = \theta(\pTveto-k_T)\,\theta(\pTveto-l_T) - \theta(\pTveto-p_T)
\end{equation} 
only involves transverse momenta and is thus the same in all regions. The function $\Delta_{ij}(k,l)$  gives the multipole expansion of the Higgs-boson on-shell constraint $(p_1+p_2-k-l)^2=m_H^2$ in the relevant momentum region. For example, one has 
\begin{equation}
   \Delta_{cc}(k,l) = \delta\big( \hat s-m_H^2 - \sqrt{\hat s}\,(k_- +l_-) \big) 
\end{equation}
in the partonic center-of-mass system. We now consider each contribution in turn.

\begin{figure}[t!]
\centering
\includegraphics[width=0.8\textwidth]{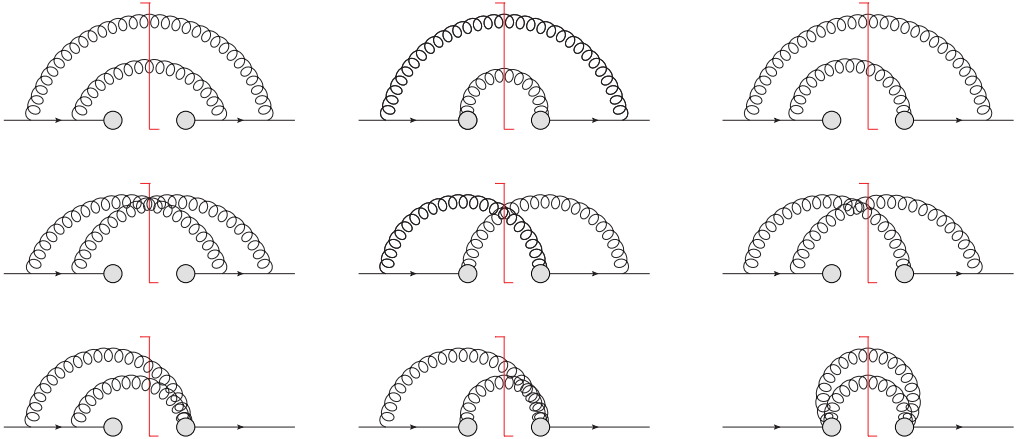}  
\caption{\label{img:collcoll}
Diagrams with color factor $C_F^2$ contributing to the squared amplitude for the independent emission of two collinear gluons. The collinear Wilson lines are represented by the gray blobs attached to the quark lines.}
\end{figure}

We begin with the ${\cal A}_{cc}$ contribution. The diagrams relevant for the $C_F^2$ color structure are shown in Figure \ref{img:collcoll}. We are only interested in the light-cone singularities of these diagrams, which result in divergences in the analytic regulator $\alpha$. Therefore only diagrams with at least one Wilson-line emission can contribute. The light-cone singularities arise from the region of the integrand in which the large minus-components of the collinear momenta tend to zero, i.e.\ when these particles become soft. In the limit where the momentum $k$ becomes soft, the $C_F^2$ part of the squared collinear amplitude takes the form
\begin{equation}\label{eq:singlesoft}
   {\cal A}_{cc}(k,l) \to \frac12\,{\cal A}_s(k)\,{\cal A}_c(l) \,,
\end{equation}
where the one-emission soft and collinear amplitudes squared are
\begin{equation}\label{eq:oneloopamps}
   {\cal A}_s(k) = \frac{16\pi C_F\alpha_s}{k_+ k_-} \,, \qquad
   {\cal A}_c(l) = 8\pi C_F\alpha_s\,\frac{2\hat s-2l_-\sqrt{\hat s}+l_-^2}{l_-l_+\hat s } \,.
\end{equation}
As we will see, only the region where both emissions become soft gives rise to a $1/\alpha$ divergence in (\ref{eq44}). In the double soft limit, the squared amplitude reduces to
\begin{equation}
   {\cal A}_{cc}(k,l) \to \frac12\,{\cal A}_s(k)\,{\cal A}_s(l) \,.
\end{equation}
With this simple form, the integration over the light-cone components becomes trivial. It has the form  
\begin{equation}\label{kpkmint}
   \int_0^\Lambda\!dk_- \int_0^\infty\!dk_+\,\delta(k^2)\,\frac{1}{k_+ k_-} 
    \left( \frac{\nu}{k_+} \right)^\alpha 
   = \frac{1}{k_T^2} \int_0^\Lambda\frac{dk_-}{k_-} \left( \frac{\nu k_-}{k_T^2} \right)^\alpha
   = \frac{1}{\alpha} \left( \frac{\nu\Lambda}{k_T^2} \right)^\alpha \frac{1}{k_T^2} \,.
\end{equation}
We have inserted an upper cutoff $\Lambda\sim m_H$ in the $k_-$ integral, since we are only interested in the divergences arising at small $k_-$. Changing variables to $p_T=k_T+l_T$ and $\xi=k_T/l_T$, and integrating over the total transverse momentum $p_T$, the integral in (\ref{eq44}) becomes
\begin{equation}\label{doublesoft}
\begin{aligned}
   \Delta\hat\sigma_{cc}(p_T) \big|_{R\rm{-indep.}}
   &= \frac12 \left( \frac{2\alpha_s C_F}{\pi} \right)^2 \frac{1}{\alpha^2}\,
    \delta(\hat s-m_H^2) \left( \frac{\Lambda^2\nu^2}{(\pTveto)^4} \right)^\alpha \\
   &\quad\times \frac{1}{2\alpha} \int_0^1\frac{d\xi}{\xi^{1+2\alpha}}
    \int_0^\pi\frac{d\Delta\phi}{\pi} \left[ (1+\xi^2+2\xi\cos\Delta\phi)^{2\alpha} - 1 \right] \\
   &= \delta(\hat s-m_H^2) \left( \frac{2\alpha_s C_F}{\pi} \right)^2
    \left( \frac{m_H^2\nu^2}{(\pTveto)^{4}} \right)^\alpha
    \frac{\zeta_3}{2}\,\frac{1}{\alpha} + {\cal O}(\alpha^0) \,,
\end{aligned}
\end{equation}
where in the last step we have replaced the cutoff scale $\Lambda$ by the Higgs mass, since we know from power counting that the full integral scales in this way. Note that, upon performing the double integral, one finds that the expression in the second line is of order $\mathcal{O}(\alpha)$, so that the final result has only a single pole in $\alpha$ even though the light-cone integrations have produced a double pole. This ${\cal O}(\alpha)$ suppression is also the reason why only the double soft limit is divergent. After subtracting the double-soft part from the total contribution $\Delta\hat\sigma_{cc}(p_T)|_{R\rm{-indep.}}$, the light-cone integrations for the single-soft contribution (\ref{eq:singlesoft}) give only rise to a single pole, and since ${\cal A}_c(q)$ has the same transverse-momentum dependence as ${\cal A}_s(q)$, the ${\cal O}(\alpha)$ suppression of the transverse-momentum integration then renders the integral finite. We conclude that only the double soft region gives rise to a divergence, so that (\ref{doublesoft}) is indeed the full result. 

Next, we consider the contribution $\Delta\hat\sigma_{\bar c\bar c}(\pTveto)$. Its structure is basically the same as above, except that the analytic regulator is now attached to the large momentum component, so that the light-cone integrations give
\begin{equation}\label{kpkmintalt}
   \int_0^\Lambda\!dk_+ \int_0^\infty\!dk_-\,\delta(k^2)\,\frac{1}{k_+ k_-} 
    \left( \frac{\nu}{k_+} \right)^\alpha 
   = \frac{1}{k_T^2} \int_0^\Lambda\frac{dk_-}{k_-} \left( \frac{\nu k_-}{k_T^2} \right)^\alpha
   = \frac{1}{\alpha} \left( \frac{\nu^2}{\Lambda^2} \right)^\alpha \frac{1}{k_T^2} \,.
\end{equation}
In contrast to (\ref{kpkmint}), the integral over transverse momentum is not affected by the regulator $\alpha$. The transverse-momentum integration associated with this term can thus be obtained by taking the $\alpha\to 0$ limit in the second line of (\ref{doublesoft}). But we have seen above that this integral is of ${\cal O}(\alpha)$, and hence it follows that
\begin{equation}
   \Delta\hat\sigma_{\bar c\bar c}(\pTveto) \big|_{R\rm{-indep.}} = 0 \,.
\end{equation}

This leaves us with the mixed contribution $\Delta\hat\sigma_{\bar c c}(\pTveto)$. Since the SCET Lagrangian does not contain any interactions coupling collinear and anti-collinear particles, the squared amplitude is a product
\begin{equation}
   {\cal A}_{\bar c c}(k,l) = {\cal A}_{\bar c}(k)\,{\cal A}_c(l) \,,
\end{equation}
where the first-order collinear amplitude squared was given in (\ref{eq:oneloopamps}) above, and the anti-collinear amplitude squared ${\cal A}_{\bar c}(k)$ is obtained from ${\cal A}_c(k)$ by interchanging $k_+$ and $k_-$. Expanding the result in the soft limit, performing the integrations over the light-cone momentum components using (\ref{kpkmint}) and (\ref{kpkmintalt}) in the two sectors, and evaluating the integrals over transverse momenta as in (\ref{doublesoft}), we get
\begin{equation}
    \Delta\hat\sigma_{\bar c c}(\pTveto) \big|_{R\rm{-indep.}}
    = - \left( \frac{2\alpha_s C_A}{\pi} \right)^2 \delta(\hat s-m_H^2)
     \left( \frac{\nu}{\pTveto} \right)^{2\alpha} \frac{\zeta_3}{2}\,\frac{1}{\alpha} 
     + {\cal O}(\alpha^0) \,.
\end{equation}

Summing the different contributions, we finally obtain 
\begin{equation}
   \Delta\hat\sigma(\pTveto) \big|_{R\rm{-indep.}}
   = \left( \frac{2\alpha_s C_A}{\pi} \right)^2 \delta(\hat s-m_H^2)\,
    \frac{\zeta_3}{2}\,\frac{1}{\alpha} \left[ \left(\frac{\nu m_H}{(\pTveto)^2}\right)^{2\alpha}
    - \left( \frac{\nu}{\pTveto} \right)^{2\alpha} \right] .
\end{equation}
The cancellation of the divergence provides a check on our computation. The resulting contribution to the anomaly coefficient derived from (\ref{Deltad2}) is 
\begin{equation}\label{resRindep}
   \Delta d_2^{\rm veto}(R) \big|_{C_F^2}^{R\rm{-indep.}} = -32\zeta_3 C_F^2 \,.
\end{equation}
Interestingly, this term exactly cancels the $\zeta_3$ term which arose in (\ref{bosontoveto}) from the Fourier integral in the expansion of the boson $q_T$ spectrum.

In the discussion above, we have exploited the fact that the light-cone singularities arise when the collinear particles become soft, and that the soft parts of the amplitudes can be factorized off. The structure of this factorization can be understood by splitting the collinear gluon field $A_c$ into a collinear and an ultrasoft gluon field, $A_c\to A_c+A_{us}$. This ultrasoft field describes collinear particles in the limit where their large light-cone momentum components become small, $k_-\sim\varepsilon m_H\ll m_H$. Its other light-cone component scales as $k_+\sim\lambda^2$, and is therefore softer than the soft mode in the factorization formula (\ref{dsigdy}). For $\varepsilon\sim\lambda^2$, this mode would be the standard ultrasoft gluon, but the relative scaling of $\varepsilon$ and $\lambda$ is not important in the following. Decoupling the ultrasoft gluon, the collinear quark field matches onto
\begin{equation}\label{eq:usoft}
   W^\dagger(x)\,\psi(x) \to W^\dagger(x)\,Y_{\bar n}^\dagger(x)\,Y_n(x)\,\psi(x) \,.
\end{equation}
The ultrasoft Wilson line $Y_{\bar n}^\dagger(x)$ arises from the substitution $A_c\to A_c+A_{us}$ in the collinear Wilson line $W^\dagger(x)$, while the second ultrasoft Wilson line arises after decoupling the ultrasoft gluons from the collinear quark field $\psi$. These ultrasoft contributions are scaleless in our regularization scheme, so we did not need to include them explicitly. But as we have shown above, we can use their structure to extract the divergences in the analytic regulator. Relation (\ref{eq:usoft}) is also the underlying reason why the cancellation of the divergences between the different sectors works: they all reduce to (ultra)soft Wilson lines in the singular limit. Since the Wilson lines arising for quarks and gluons only differ in their color representation, we can obtain the gluon result from the quark result computed above by replacing $C_F\to C_A$.

We now have computed all the ingredients required to present the complete result for the two-loop anomaly coefficient $d_2^{\rm veto}(R)$. Combining (\ref{bosontoveto}) and (\ref{Deltadd2}), we obtain
\begin{equation}\label{d2vetoR}
   d_2^{\rm veto}(R) = d_2^B - 32 C_B\,f_B(R) \,,
\end{equation}
with
\begin{equation}\label{fBRres}
\begin{aligned}
   f_B(R) 
   &= C_A\,\Big( c_L^A \ln R + c_0^A + c_2^A R^2 + c_4^A R^4 + \dots \Big) 
    + C_B \left( - \frac{\pi^2 R^2}{12} + \frac{R^4}{16} \right) \\
   &\quad\mbox{}+ T_F n_f \,\Big( c_L^f \ln R + c_0^f + c_2^f R^2 + c_4^f R^4 + \dots \Big) \,.
\end{aligned}
\end{equation}
For the Higgs case, with $C_B=C_A$, this reproduces the numerical result given in (\ref{fR}).

\section{Two-loop beam functions and fixed-order matching}
\label{sec:beamfuns}

The one remaining unknown two-loop ingredient to the factorization theorem (\ref{sigfinal}) is the two-loop beam function $\bar B_g(\xi,\pTveto)$ defined in (\ref{colanom}). In (\ref{pdfmatch}) we have matched this function onto standard PDFs, and we have then presented the one-loop expressions for the kernel functions $\bar I_{g\leftarrow i}$. For our analysis we will extract the two-loop contributions to $\bar B_g$ numerically. At the same time, we will match our resummed expression for the jet-veto cross section with the corresponding fixed-order expression at ${\cal O}(\alpha_s^2)$. In this way, we extract terms that are power-suppressed in the small ratio $\pTveto/m_H$. Once this is done, our result not only resums the large logarithms of $m_H/\pTveto$ at N$^3$LL$_{\rm p}$ order, but it also accounts for all two-loop corrections. 

At fixed order in perturbation theory, the two-loop result for the Higgs cross section with a jet veto can be obtained by running the codes {\sc FeHiP} \cite{Anastasiou:2005qj} or {\sc HNNLO} \cite{Catani:2007vq,Grazzini:2008tf}. These Monte-Carlo programs compute the production cross section at ${\cal O}(\alpha_s^2)$, with arbitrary cuts on the final state. In the following, we use {\sc HNNLO} with MSTW2008NNLO PDFs \cite{Martin:2009iq} and $\alpha_s(m_Z)=0.1171$. In order to extract the product of the two beam functions with two-loop precision, we compute the cross section integrated over rapidity and divide it by the perturbative expansion for the hard function $\bar H$ defined in (\ref{Hbardef}). This yields the reduced cross section
\begin{equation}\label{reducedXS}
   \bar\sigma(\pTveto) = \frac{\sigma(\pTveto)}{\bar H(m_t,m_H,\pTveto)} 
   \equiv \bar\sigma_\infty(\pTveto) + \Delta\bar\sigma(\pTveto) \,,
\end{equation}
with
\begin{equation}\label{sigreduced}
   \bar\sigma_\infty(\pTveto)
   = \sigma_0(\pTveto)\,\int_{-y_{\rm max}}^{y_{\rm max}}\!dy\,
    \bar B_g(\tau e^y,\pTveto)\,\bar B_g(\tau e^{-y},\pTveto) \,,
\end{equation}
where $\tau=m_H/\sqrt{s}$ and $y_{\rm max}=\ln(1/\tau)$. The quantity $\bar\sigma_\infty$ contains the leading-power contribution and is proportional to the convolution of the two beam functions. The remainder $\Delta\bar\sigma={\cal O}(\pTveto/m_H)$ in (\ref{reducedXS}) contains the power corrections to the reduced cross section. The rationale for considering the reduced cross section is that, in the factorization formula (\ref{sigfinal}), all large logarithms are resummed in the RG-invariant hard function $\bar H$ (provided we choose $\mu\sim\pTveto$). The reduced cross section obtained when $\bar H$ is factored out has a well-behaved perturbative expansion, and it can thus be extracted from numerical fixed-order codes. 

We now exploit the fact that the leading-power reduced cross section $\bar\sigma_\infty$ depends on $m_H$ only through the ratio $m_H/\sqrt{s}$, which enters in the arguments of the beam functions and in $\sigma_0(\pTveto)$. If we compute the reduced cross section for a very large value of $m_H$, keeping the ratio $m_H/\sqrt{s}$ fixed at its physical value, the power corrections will become negligibly small and we directly obtain the quantity $\bar\sigma_\infty$, and from it the two-loop beam functions. Repeating the analysis with the physical value $m_H=125$\,GeV, we are then able to extract the power-suppressed contribution $\Delta\bar\sigma$. In practice, we run the program {\sc HNNLO} at a fixed value of $\mu=\mu_f=\mu_r$, once with the physical values $m_H=125$\,GeV and $\sqrt{s}=8$\,TeV, and a second time with the larger values $m_H=500$\,GeV and $\sqrt{s}=32$\,TeV. The latter value for the Higgs mass is sufficiently large to ensure that power-suppressed terms are very small in the range of $\pTveto$ values we are considering. To very good approximation, the power corrections can then be obtained from the difference 
\begin{equation}
   \Delta\bar\sigma(\pTveto) \simeq \bar\sigma(\pTveto) \big|_{m_H=125\,{\rm GeV}}
    - \bar\sigma(\pTveto) \big|_{m_H=500\,{\rm GeV}} \,.
\end{equation}

\begin{figure}
\begin{center}
\includegraphics[width=0.33\textwidth]{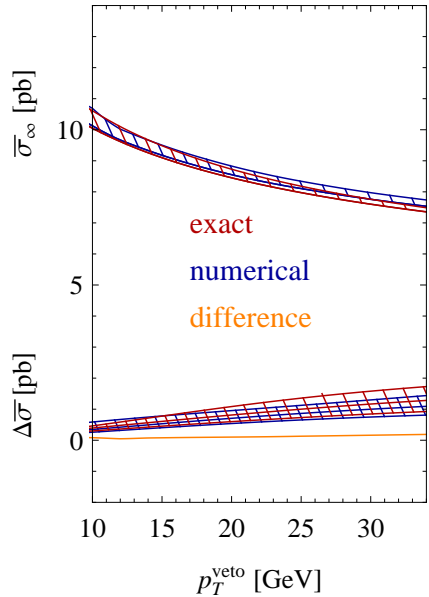}
\vspace{-0.3cm}
\caption{\label{fig:NLOcheck}
Comparison of the exact NLO results for the reduced cross section $\bar\sigma_\infty$ (upper red band) and its power corrections $\Delta\bar\sigma$ (lower red band) with the corresponding numerical results extracted using the procedure outlined in the text (upper and lower blue bands). The orange band shows the difference between the exact and numerical results.}
\end{center}
\end{figure}

As a validation, we have performed the numerical extraction of the beam functions and power corrections also at NLO, where the expression for $\bar B_g(\xi,\pTveto)$ is known analytically, see (\ref{pdfmatch}). The two upper bands in Figure~\ref{fig:NLOcheck} show the leading-power reduced cross section $\bar\sigma_\infty$, while the two lower bands show the power-suppressed contribution $\Delta\bar\sigma$. In all cases we show NLO bands obtained by varying the factorization scale in the region $\pTveto/2<\mu<2\pTveto$. The blue bands show the numerical result extracted from the procedure just described, while the red bands give the exact result, obtained by using the analytic expressions (\ref{Ires}) for the calculation of the beam functions. We observe that the numerical method reproduces the analytical results with good accuracy. The small difference is shown by the very narrow orange band in the figure. This band is equal to the power corrections at $m_H=500$\,GeV, which are very small but non-zero. In our final matched results, we add back the power-suppressed terms to $\bar{\sigma}(\pTveto)$, so that the small residual power corrections remaining at $m_H=500$\,GeV do not change our predictions for the cross section. We separate out the power corrections in order to assess their relative size and to be able to vary the scale $\mu$ independently for $\bar\sigma_\infty$ and $\Delta\bar\sigma$. The small scale uncertainty of the fixed-order cross section is known to be due to a cancellation of different types of large corrections. In order to avoid such accidental cancellations, we separate the different parts of the calculation and vary their scales independently.

\begin{figure}
\begin{center}
\begin{tabular}{ccc}
\includegraphics[width=0.299\textwidth]{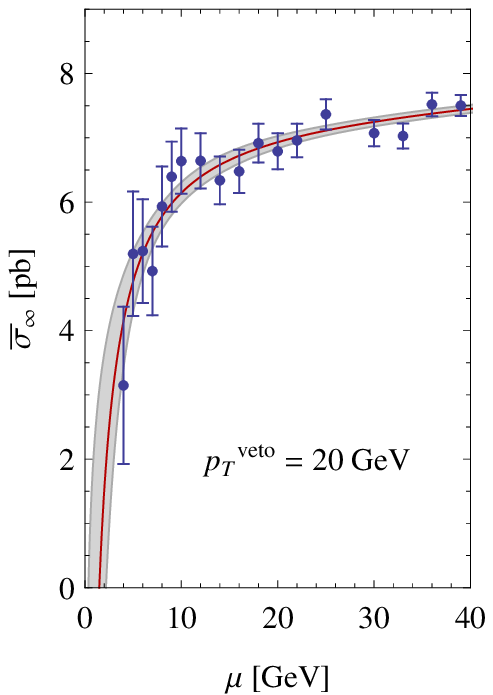}& \includegraphics[width=0.312\textwidth]{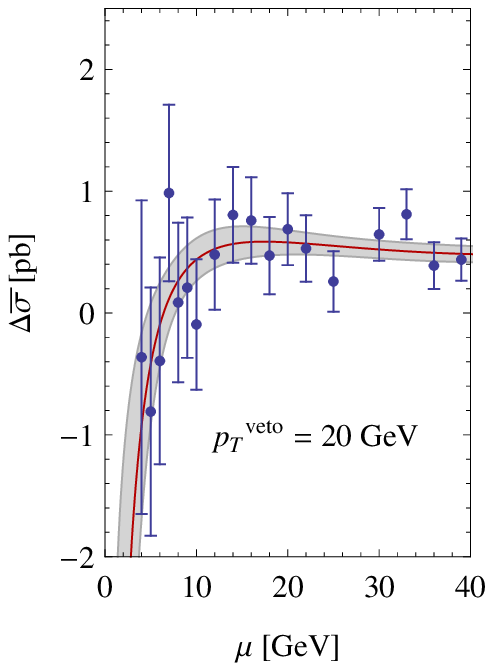} & \includegraphics[width=0.307\textwidth]{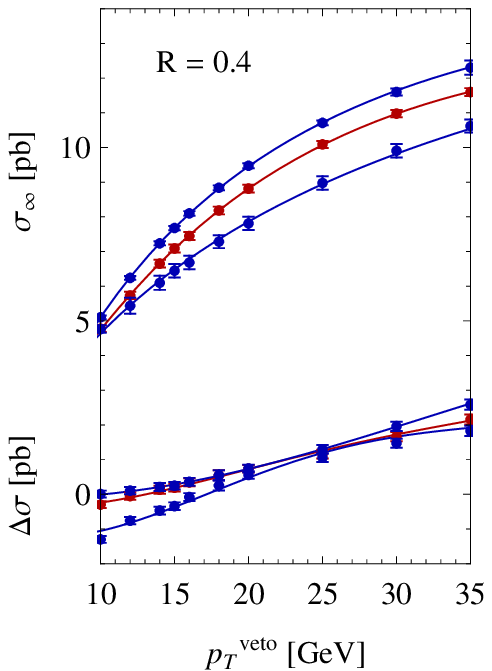} 
\end{tabular}
\vspace{-0.3cm}
\caption{\label{fig:Fitting} 
Numerical results for the reduced cross section $\bar\sigma_\infty$ and its power corrections $\Delta\bar\sigma$ at NNLO. The first two plots show the results of a fit to the $\mu$ dependence of these quantities at fixed $\pTveto=20$\,GeV and $R=0.4$. The right plot shows our fit to the $\pTveto$ dependence of the cross section $\sigma_\infty=\bar H\bar\sigma_\infty$ and its power corrections $\Delta\sigma=\bar H\Delta\bar\sigma$ (obtained using the function $\bar H$ evaluated at its default scale $\mu=\pTveto$), indicating their default values (red lines) and scale variation (blue lines).}
\end{center}
\end{figure}

At NNLO the numerics become more challenging, in particular at the high value $m_H=500\,{\rm GeV}$. In the left two plots in Figure~\ref{fig:Fitting}, we show our numerical results for the leading-power cross section $\bar\sigma_\infty(\pTveto)$ (left) as well as for the power corrections $\Delta\bar\sigma(\pTveto)$ (center) for $\pTveto=20$\,GeV and $R=0.4$, as a function of the factorization scale $\mu$. We generate a grid of 24 different $\mu$ values and 5 different choices of the jet radius $R$. For each parameter pair, we perform 20 independent runs of the {\sc HNNLO} program, each producing $3\cdot 10^8$ events. Every run generates a histogram for $\bar\sigma(\pTveto)$ with the selected parameter values and takes approximately 10 hours to complete, so that the total computing time would amount to 2000 days on a single processor core. Despite the large number of events, the statistical uncertainties on the extracted values in Figure~\ref{fig:Fitting} are not completely negligible. To determine the default value and the scale variation at a given value of $\pTveto$, we fit a third-order polynomial in $\ln\mu$ to the numerical data. The resulting fit functions, together with their uncertainties, are shown in the left two panels of Figure~\ref{fig:Fitting}. In both cases, the quality of the fit is excellent ($\chi^2/{\rm dof}\approx 0.8$). From the fit in $\mu$, we extract the default value for the cross section and the upper and lower edges of the scale-variation band, for each value of $\pTveto$. In a last step, we first multiply by the prefactor $\bar H$ evaluated at its default scale $\mu=\pTveto$ and then fit a third-order polynomial in $\ln\pTveto$ to the leading-power cross-section results, and a fourth-order polynomial in $\pTveto$ to the power corrections. We do not include a constant term in the fit to $\Delta\bar\sigma$, since the power corrections must vanish for $\pTveto\to 0$. The fitted curves are shown in the third plot in Figure~\ref{fig:Fitting}. Once again, the upper curves show the leading-power cross section $\bar\sigma_\infty(\pTveto)$ together with its scale-uncertainty band, while the lower ones show the corresponding results for the power corrections $\Delta\bar\sigma(\pTveto)$. The fact that the scale variation at NNLO turns out to be larger than at NLO can be traced back to the presence of rather large, $R$-dependent two-loop corrections in the beam functions. This will be discussed in more detail in Section~\ref{sec:numerics}. We have also used other forms of fit functions and find compatible results. However, employing too many fit parameters would cause the fit to follow the statistical fluctuations of the numerical results. As a further cross check, we have also computed the $\pTveto$ dependence using the {\sc MCFM} code \cite{MCFM} instead of {\sc HNNLO}, finding results consistent with the ones presented here.

\begin{figure}
\begin{center}
\begin{tabular}{c}
\includegraphics[width=0.7\textwidth]{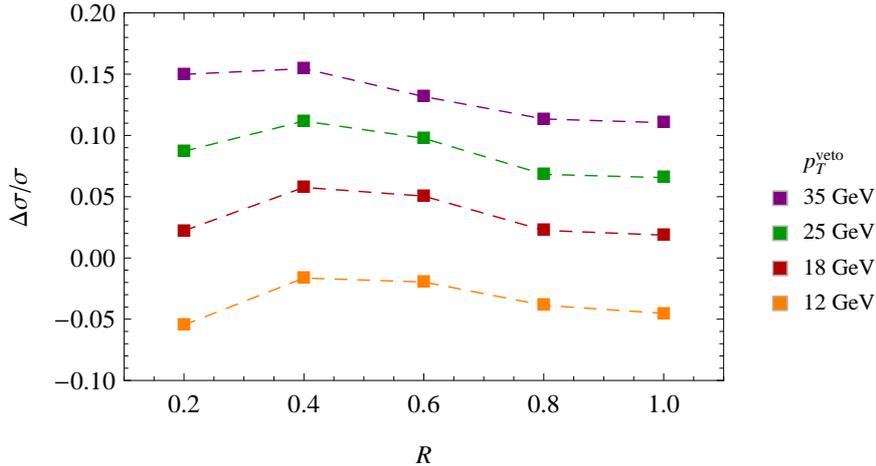} 
\end{tabular}
\vspace{-0.3cm}
\caption{\label{fig:RdependencePowerCorrections}
Jet-radius dependence of the relative size of the power corrections, $\Delta\sigma/\sigma$, for different values of the veto scale $\pTveto$.}
\end{center}
\end{figure}

It is interesting to look at the dependence of the power corrections on the jet-radius parameter $R$. From (\ref{eq:phasespace}), one would naively expect that the power corrections can be enhanced by factors of $e^R$, as mentioned near the end of Section~\ref{sec:multipole}. However, numerically we see no evidence for such an effect. Indeed, as can be seen from Figure~\ref{fig:RdependencePowerCorrections}, we find a very moderate dependence on the jet radius. The relative size of the power corrections, $\Delta\sigma(\pTveto)/\sigma(\pTveto)=\Delta\bar\sigma(\pTveto)/\bar\sigma(\pTveto)$, turns out to be almost independent of $R$ in the range $0.2<R<1$.

\section{Numerical predictions for the LHC}
\label{sec:numerics}

\begin{figure}
\begin{center}
\begin{tabular}{c}
\includegraphics[width=0.99\textwidth]{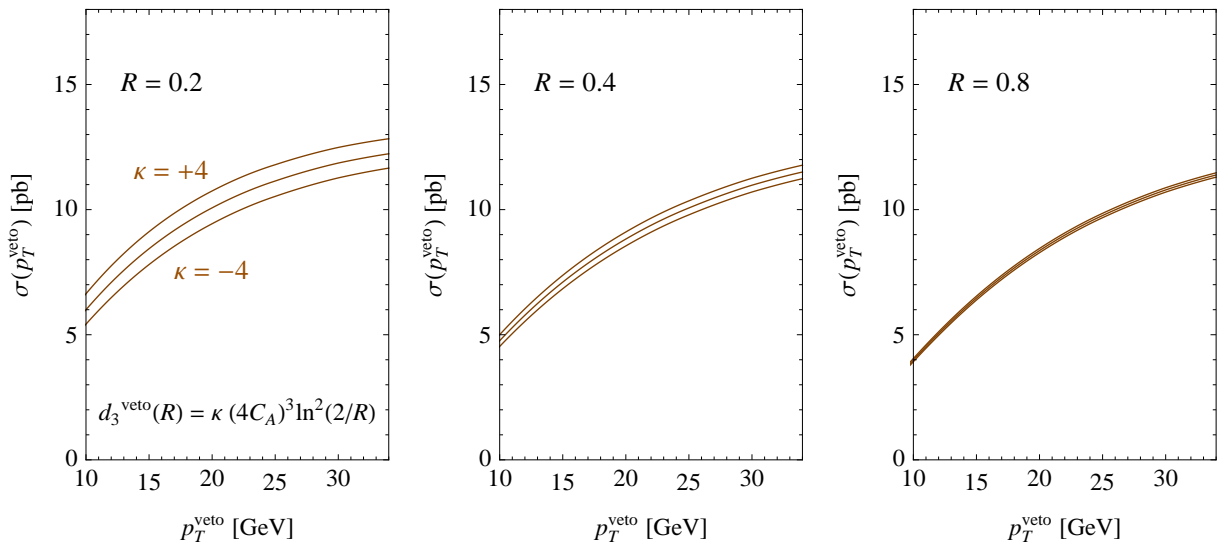}
\end{tabular}
\vspace{-0.5cm}
\caption{\label{fig:d3variation} 
Uncertainty in the jet-veto cross section due the variation of the three-loop anomaly coefficient $d_3^{\rm veto}(R)$ for three different values of the jet radius.}
\end{center}
\end{figure}

We are now in a position to present our final results for the jet-veto cross section and the veto efficiency for Higgs-boson production in gluon fusion at the LHC. In order to obtain the highest possible accuracy at present, we combine resummed results at N$^3$LL$_{\rm p}$ order with fixed-order results at NNLO in perturbation theory. The only missing ingredients for a complete resummation with N$^3$LL accuracy are the four-loop coefficient $\Gamma_3^A$ of the cusp anomalous dimension and the three-loop coefficient $d_3^{\rm veto}(R)$ in the anomaly exponent $F_{gg}$ in (\ref{Fhexpansions}). Both quantities enter via the RG-invariant hard function defined in (\ref{Hbardef}). For the four-loop cusp anomalous dimension, we use the Pad\'e approximation
\begin{equation}
   \Gamma_3^A \big|_{\mbox{\scriptsize Pad\'e}} = \frac{(\Gamma_2^A)^2}{\Gamma_1^A} = 3494.4 \,,
\end{equation}
valid for $n_f=5$. A corresponding estimate works very well one order lower, where one has $\Gamma_2^A=538.2$ and $(\Gamma_1^A)^2/\Gamma_0^A=572.7$. The largest effect of $\Gamma_3^A$ occurs at low $\pTveto$ values. However, even at the very low value $\pTveto=10$\,GeV, switching off the four-loop cusp anomalous dimension would increase the cross section by only 0.1\%, so that the uncertainty associated with $\Gamma_3^A$ is negligibly small. The contribution of the unknown three-loop anomaly coefficient $d_3^{\rm veto}(R)$ to the cross section is of the form $(\alpha_s/\pi)^3\ln(\pTveto/m_H)$. Generically, we would expect this type of contribution to be small in the range of $\pTveto$ values we consider, since the logarithm $\ln(\pTveto/m_H)$ is not large enough to fully compensate the suppression by a factor of $\alpha_s/\pi$. However, we have seen in Section~\ref{sec:2loop} that the anomaly coefficient is enhanced at small $R$ by factors of $\ln R$. The leading-color part of the two-loop coefficient can be well approximated as $d_2^{\rm veto}(R)\approx 2\,(4C_A)^2 \ln(2/R)$. Motivated by this, we will estimate the quantity $d_3^{\rm veto}(R)$ as
\begin{equation}\label{d3estimate}
   d_3^{\rm veto}(R) = \kappa \left( 4C_A \right)^3 \ln^2\!\frac{2}{R} \,,
\end{equation}
and vary the overall prefactor in the range $-4<\kappa<4$. The result of this variation on the cross section is shown in Figure~\ref{fig:d3variation}. The above ansatz encodes the correct logarithmic scaling at small $R$, and we believe it provides a generous estimate for all $R$ values considered in our work. Even at $R=1$ our estimate for $d_3^{\rm veto}(R)$ is still more than six times larger than the three-loop cusp anomalous dimension $\Gamma_2^A$. Nevertheless, the resulting effect is seen to be very small for larger values of $R$. Also for smaller values, such as $R=0.4$, the associated uncertainty is lower than the scale uncertainty. While a full computation of $d_3^{\rm veto}(R)$ looks difficult, we believe that a determination of the coefficient of the leading logarithm should be feasible. The double logarithm arises from diagrams with three collinear emissions, which involve two propagators that are nearly on-shell. 

\subsection{Scale uncertainties}

We now proceed to explore the perturbative uncertainties in the resummed predictions for the jet-veto cross section, as estimated by scale variations. We obtain predictions for the cross section integrated over rapidity by using the RG-improved result for the hard function $\bar H$ in (\ref{Hbardef}) and multiplying it with the reduced cross section in (\ref{reducedXS}), which we have extracted with two-loop accuracy and including power corrections. Since the Sudakov logarithms exponentiate, it is natural to perform the perturbative expansion of the hard function in the exponent, i.e.\ to expand $\ln \bar H$ instead of $\bar H$ itself. For this reason, we do not perform an additional expansion after multiplying the reduced cross section by $\bar H$. To estimate the residual scale uncertainties of our predictions, we independently vary the hard matching scales $\mu_t$ and $\mu_h$, at which the Wilson coefficients $C_t$ and $C_S$ in (\ref{dsigdy}) are calculated (for details see \cite{Ahrens:2008nc}), as well as the factorization scale $\mu$, by factors of 2 about their default values $\mu_t=m_t$, $\mu_h^2=-m_H^2$, and $\mu=\pTveto$. We then obtain individual error estimates for the hard function $\bar H$ and for the reduced cross section $\bar\sigma_\infty$ and its power corrections $\Delta\bar\sigma$. The error associated with the hard function also includes the uncertainty arising due to the unknown value of $d_3^{\rm veto}(R)$, which we estimate by scanning $\kappa$ over the interval between $-4$ and 4. Beyond NLL order, the sensitivity to variations of the hard scales $\mu_t$ and $\mu_h$ is so small that one can safely neglect it compared to the effect of the $\mu$ variation. For instance, at $\pTveto=20$\,GeV the $\mu_h$ variation is $\pm 0.3\%$ and the $\mu_t$ variation $^{+0.1}_{-0.2}\%$ at N$^3$LL level ($\pm 1\%$ in both cases at NNLL order). Since the quantities $\bar H$ and $\bar\sigma$ are RG invariant, it seems reasonable to assume that their residual scale uncertainties are uncorrelated. We therefore combine the errors in $\bar H$, $\bar\sigma_\infty$, and $\Delta\bar\sigma$ in quadrature to obtain our final error estimates. 

In addition to the matching and factorization scales, one can also consider a variation of the logarithms associated with the collinear anomaly. This can be done by rewriting the anomaly factor in (\ref{Hbardef}) in the form
\begin{equation}
   \bigg( \frac{m_H}{\pTveto} \bigg)^{-2F_{gg}(\pTveto,\mu)} 
   = \bigg( \frac{m_H}{\nu} \bigg)^{-2F_{gg}(\pTveto,\mu)} 
    \bigg( \frac{\nu}{\pTveto} \bigg)^{-2F_{gg}(\pTveto,\mu)} \,.
\end{equation}
For $\nu\sim\pTveto$, the second factor on the right-hand can be expanded in fixed-order perturbation theory, after which some higher-order $\nu$ dependence is left over. For example, at NNLL order the one-loop expression for $F_{gg}$ in (\ref{Fhexpansions}) is sufficient for the second factor, while the two-loop expression is needed for the first one beause of the large logarithm. The variation from changing $\nu$ by a factor of 2 about the default value $\nu=\pTveto$ is $\pm 10\%$ at NNLL order, while it vanishes by definition at N$^3$LL order if $d_3^{\rm veto}(R)=0$ and $\mu=\pTveto$, assuming the expansion is performed for the logarithm of $\bar H(m_t,m_H,\pTveto)$, as we do. If instead $\bar H(m_t,m_H,\pTveto)$ itself is expanded, then the variation is $\pm 3\%$. The type of scale variation considered here can be formalized in an RG framework \cite{Chiu:2011qc,Chiu:2012ir}, in which the change in $\nu$ reshuffles contributions between the soft and collinear functions. However, in contrast to the standard RG, there is no physical coupling constant involved in the running, since the different contributions live at the same virtuality. Furthermore, the individual contributions are strongly scheme dependent. With our regulator, all perturbative corrections to the soft function vanish, while the regulator put forward in \cite{Chiu:2012ir} leads to a non-zero soft function. For these reasons, we do not believe that the $\nu$ variation provides much insight into the size of higher-order corrections, and we therefore do not include it in our error budget.

\begin{figure}[t!]
\begin{center}
\includegraphics[width=0.99\textwidth]{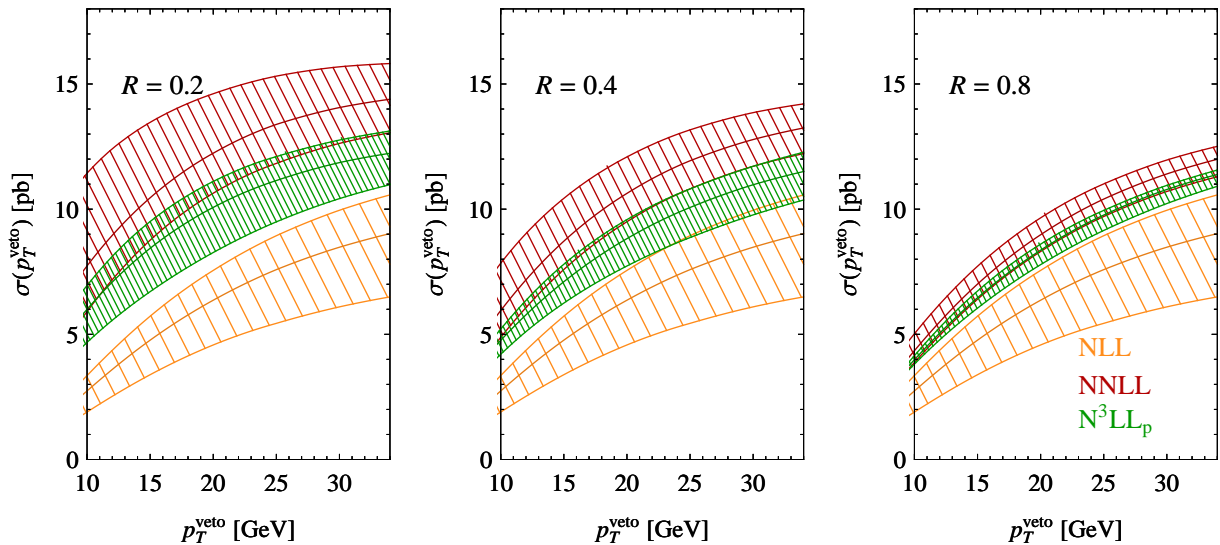}
\vspace{-0.5cm}
\caption{\label{fig:resummed} 
Resummed predictions for the leading-power jet-veto cross section at NLL (orange), NNLL (red), and N$^3$LL$_{\rm p}$ order (green).}
\end{center}
\end{figure}

\begin{figure}[t!]
\begin{center}
\includegraphics[width=0.5\textwidth]{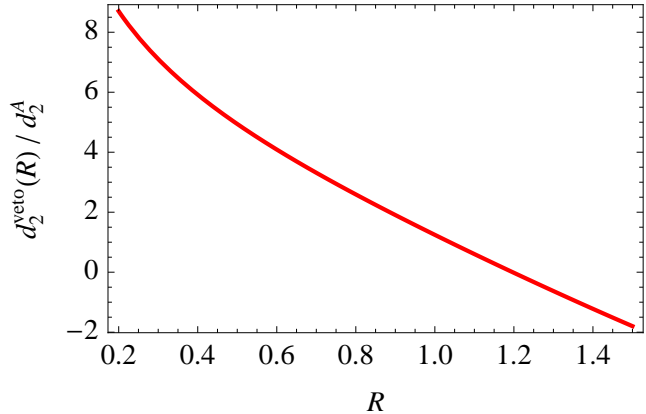}
\vspace{0.0cm}
\caption{\label{fig:plR}
Dependence of the two-loop anomaly coefficient $d_2^{\rm veto}(R)$ (in units of the coefficient $d_2^A$) on the jet radius.}
\end{center}
\end{figure}

Figure~\ref{fig:resummed} shows our predictions for the leading-power cross section for three different values of the jet-radius parameter $R$. The colored bands refer to the predictions obtained at NLL, NNLL, and N$^3$LL$_{\rm p}$ order. Consider first the right-most panel, which corresponds to the relatively large value $R=0.8$. In this case we observe a reduction of the scale uncertainties as we increase the accuracy of the resummation. While the NLL and NNLL bands do not quite overlap, they are at least near each other. The N$^3$LL$_{\rm p}$ band overlaps with the NNLL band and counteracts to some extent the large enhancement seen at NNLL order. All in all, it appears that the impact of higher-order effects is roughly in accordance with the error estimates from lower-order results, suggesting that the perturbative series is reasonably well behaved. Unfortunately, the quality of the expansion deteriorates as one lowers the jet radius $R$. The size of the corrections and the uncertainties obtained at NNLL and N$^3$LL$_{\rm p}$ order both increase with decreasing $R$. For $R=0.2$, the NNLL band is as broad as (or even broader than) the NLL band, and there is a rather substantial gap between them. The origin of the large scale dependence of the NNLL order bands at small $R$ can be traced back to the behavior of the two-loop anomaly coefficient $d_2^{\rm veto}(R)$ given in (\ref{notCasi}), which is plotted in Figure~\ref{fig:plR} in units of the coefficient $d_2^A$ appearing in the resummation formula for the transverse-momentum distribution of Higgs bosons at low $q_T\ll m_H$ \cite{Becher:2012yn}. Whereas $d_2^{\rm veto}(R)/d_2^A$ is of modest size for $R\gtrsim 0.8$, this ratio quickly increases as $R$ decreases, and it reaches a very large value $d_2^{\rm veto}(R)/d_2^A\approx 8.7$ for $R=0.2$. The origin of this effect can be understood from the presence of the $\ln R$ term in the expression for the function $f(R)$ in (\ref{fR}), which becomes large for such small values of the jet radius. Note that the $d_2^{\rm veto}(R)$ term first appears at NNLL order, and that the $\mu$ dependence of the running coupling in the anomaly term 
\begin{equation}\label{above}
   \exp\left[ - \frac{d_2^{\rm veto}(R)}{8} \left( \frac{\alpha_s(\mu)}{\pi} \right)^2 
    \ln\frac{m_H}{\pTveto} \right]
   \approx \exp\left[ 1.21\,\frac{d_2^{\rm veto}(R)}{d_2^A}\,\alpha_s^2(\mu)\,
    \ln\frac{m_H}{\pTveto} \right]
\end{equation}
contained in the hard function $\bar H$ in (\ref{Hbardef}) only gets compensated at N$^3$LL order. For $\pTveto=25$\,GeV and $R=0.2$, the exponent approximately equals $17\alpha_s^2(\mu)$. Since the NLL band completely misses this genuine source of large scale dependence, it underestimates the perturbative uncertainties for small $R$. To reduce the scale variations of the NNLL band, it is necessary to perform the resummation at N$^3$LL$_{\rm p}$ order, as we do in the present work. The fact that the green bands in Figure~\ref{fig:resummed} are narrower than at NNLL order and fall between the NLL and NNLL bands gives us confidence that at N$^3$LL$_{\rm p}$ order, and for $R\ge 0.4$ not too small, one captures the main corrections and obtains reliable predictions and error estimates.

\begin{figure}[t!]
\begin{center}
\includegraphics[width=0.99\textwidth]{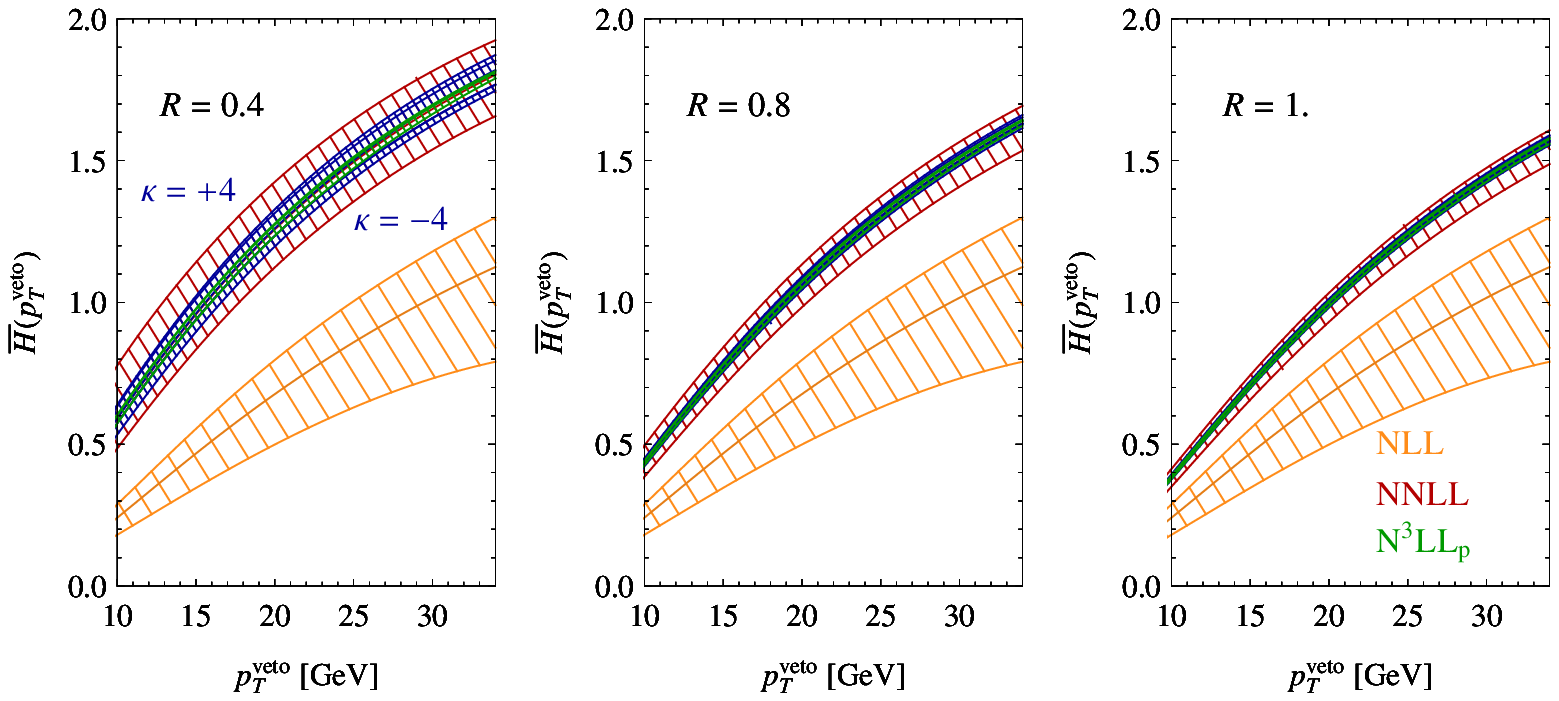}
\includegraphics[width=0.99\textwidth]{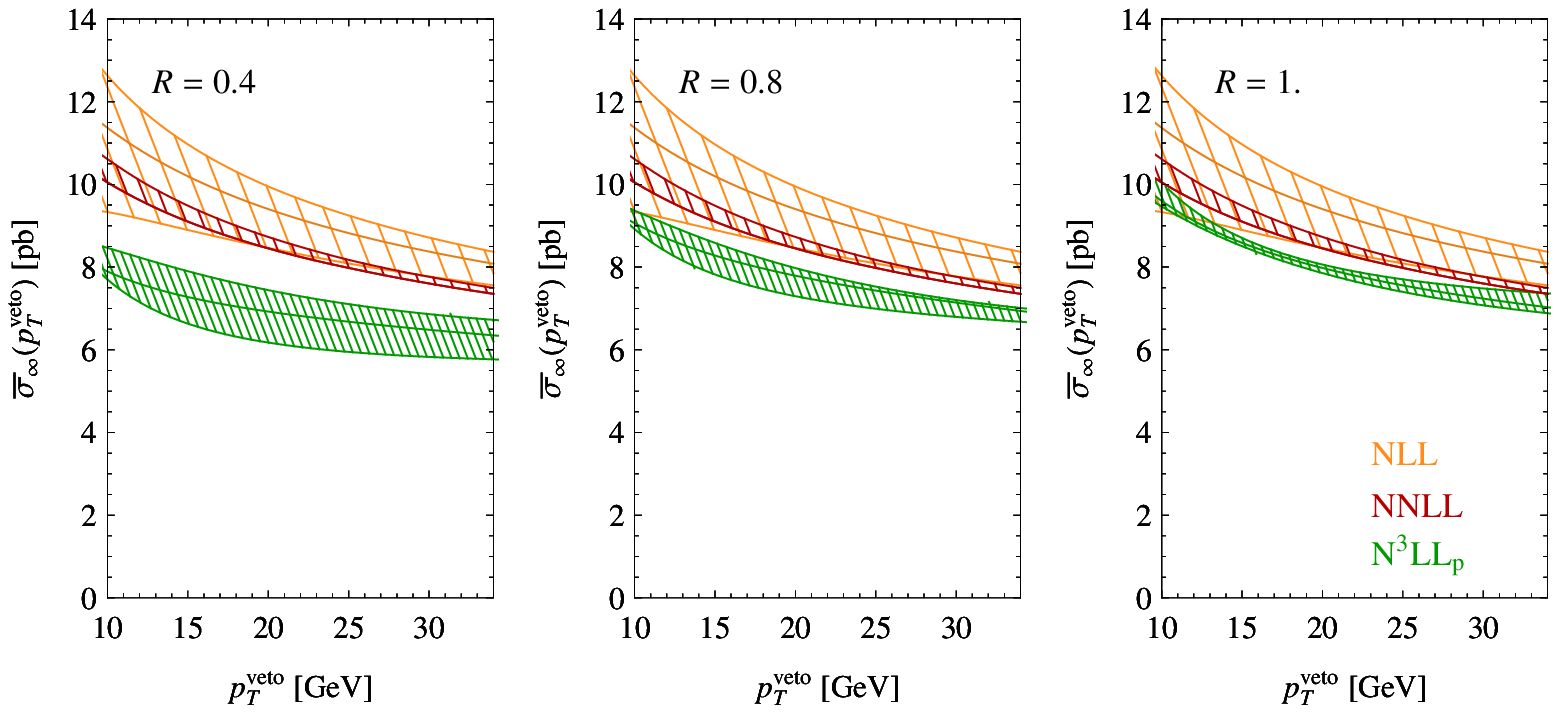} 
\vspace{-0.5cm}
\caption{\label{fig:ScaleVar}
Individual residual scale variations of the hard function $\bar H$ (top three plots) and the residual cross section $\bar\sigma_\infty$ (bottom three plots) for different $R$ values. The dark blue bands in the top panels are obtained by also including the variation of the unknown three-loop coefficient $d_3^{\rm veto}(R)$, using the estimate (\ref{d3estimate}).}
\end{center}
\end{figure}

In order to substantiate this claim, we study the scale variations of the different ingredients in the factorization formula (\ref{sigfinal}) separately. The top panels in Figure~\ref{fig:ScaleVar} show the residual scale dependence of the RG-invariant hard function $\bar H$ at different orders in perturbation theory. We observe a very large correction when going from NLL to NNLL order, whereas the impact of yet higher-order corrections is seen to be small. Indeed, comparing with Figure~\ref{fig:resummed}, we see that this dependence of the hard function explains the scale variations of the cross section shown in Figure~\ref{fig:resummed} at NLL and NNLL order. At N$^3$LL$_{\rm p}$ order, the large scale uncertainty related to the $d_2^{\rm veto}(R)$ term in (\ref{above}) gets compensated by including the three-loop terms in the anomaly exponent. From that point on, the remaining scale variation of the hard function is very small, even if we simultaneously vary the coefficient $d_3^{\rm veto}(R)$ according to our estimate (\ref{d3estimate}). This latter effect is illustrated by the difference between the green and blue bands in the plots. The bottom panels in Figure~\ref{fig:ScaleVar} show the scale variation of the leading-power reduced cross section $\bar\sigma_\infty(\pTveto)$ in (\ref{reducedXS}). Once again large $R$-dependent two-loop corrections arise, which increase with decreasing $R$. However, in the present case these corrections are contained in the beam functions and count as N$^3$LL order effects, because the reduced cross section does not contain large logarithms of $\pTveto/m_H$. As a result, while the one-loop corrections are seen to be small, at two-loop order the reduced cross section receives large negative corrections, whose size is not anticipated by the small scale dependence of the one-loop result. In addition, also the scale uncertainties increase when these corrections are included, especially at low $R$ values. Indeed, it is the residual scale dependence of the beam functions at two-loop order which dominates the scale uncertainty in our final result for the cross section (cf.\ Figure~\ref{fig:resummed}). The figures show that, once again, these large two-loop effects strongly increase with decreasing $R$. For the anomaly coefficient, we had found that the remaining higher-order corrections are moderate once the leading $\ln R$-enhanced terms are in place, and we believe that the same is true for the beam functions. In order to check that this is indeed the case, it would be necessary to compute the leading three-loop corrections to the beam functions in the limit of very small $R$ -- a task that is well beyond the scope of the present work.  

\subsection{Predictions for the jet-veto cross section}

\begin{figure}[t!]
\begin{center}
\includegraphics[width=0.99\textwidth]{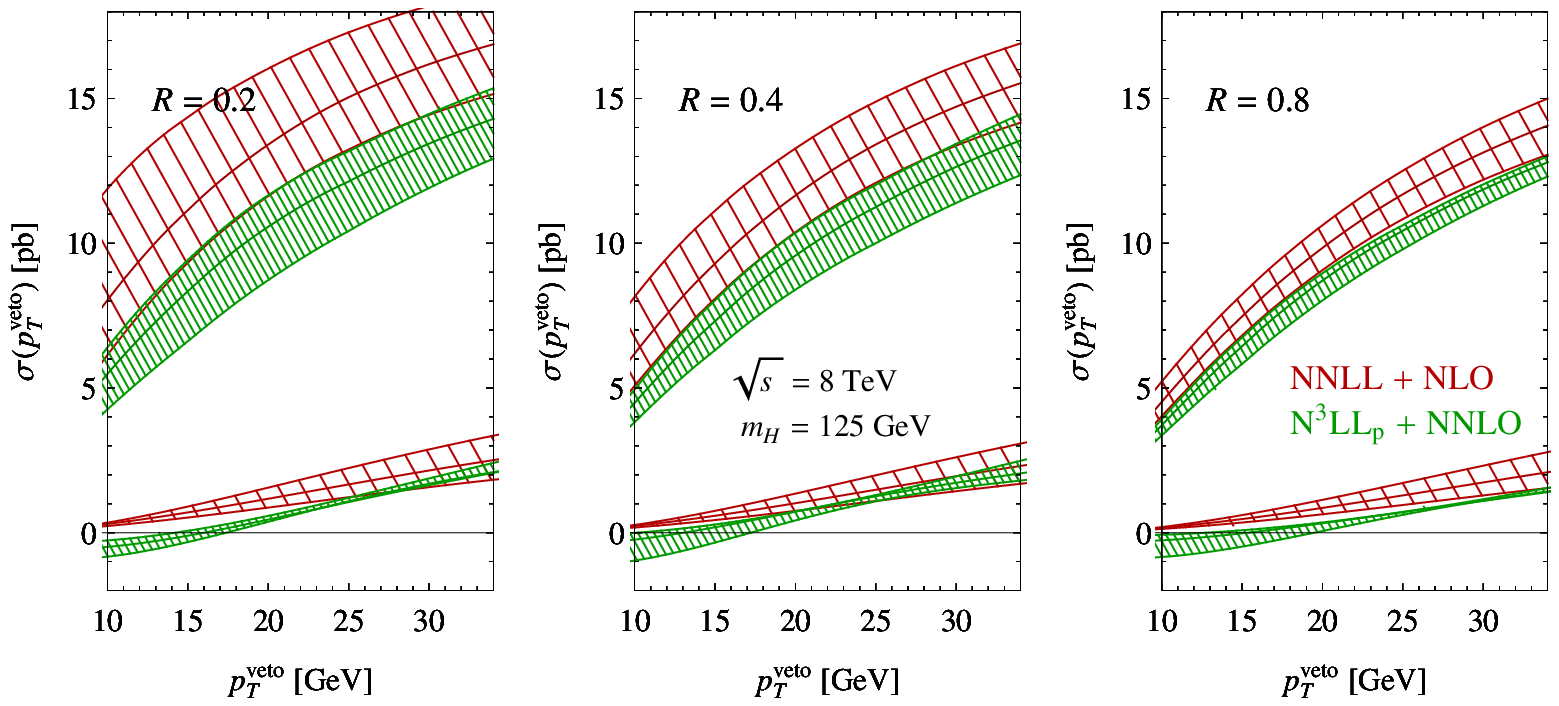}
\vspace{-0.5cm}
\caption{\label{fig:resummed2}
Resummed and matched results for the jet-veto cross section for Higgs production at the LHC. The green bands show our best predictions at N$^3$LL$_{\rm p}$+NNLO, while the red bands show for comparison the results obtained at NNLL+NLO. The uncertainty band is obtained by simultaneously varying $\pTveto/2<\mu<2\pTveto$ and the coefficient $d_3^{\rm veto}(R)$ according to the estimate (\ref{d3estimate}).}
\end{center}
\end{figure}

In the last step, we now add the power-suppressed corrections to our resummed predictions for the jet-veto cross section, thereby extending the accuracy of our results to N$^3$LL$_{\rm p}$+NNLO. Our final predictions for the cross section are depicted in Figure~\ref{fig:resummed2}. The red and green hatched bands show the cross section and its uncertainty at NNLL+NLO and N$^3$LL$_{\rm p}$+NNLO accuracy, respectively. Also shown in the bottom half of each panel are the contributions from power-suppressed effects, which are seen to remain small even for the largest $\pTveto$ values considered here. The bands in the figure account for the scale variations and, at N$^3$LL$_{\rm p}$ order, include the uncertainty due to the unknown coefficient $d_3^{\rm veto}(R)$. While the uncertainty bands obtained at different orders do not quite overlap, they lie close to each other. Given the discussion above, this is the best we could have hoped for. The scale uncertainties increase significantly as $R$ is lowered to smaller values. To have good theoretical control, one should either try to increase the value of $R$ above 0.4 in the experimental analyses, or to resum the $\ln R$-enhanced terms in the cross section. For a larger jet-radius parameter $R=0.8$ and an intermediate jet-veto scale $\pTveto= 25$\,GeV, the cross section is predicted with an accuracy of about $\pm 6\%$. For smaller $R=0.4$, the uncertainties increase to about $\pm 11\%$.  

\begin{figure}
\begin{center}
\includegraphics[width=0.99\textwidth]{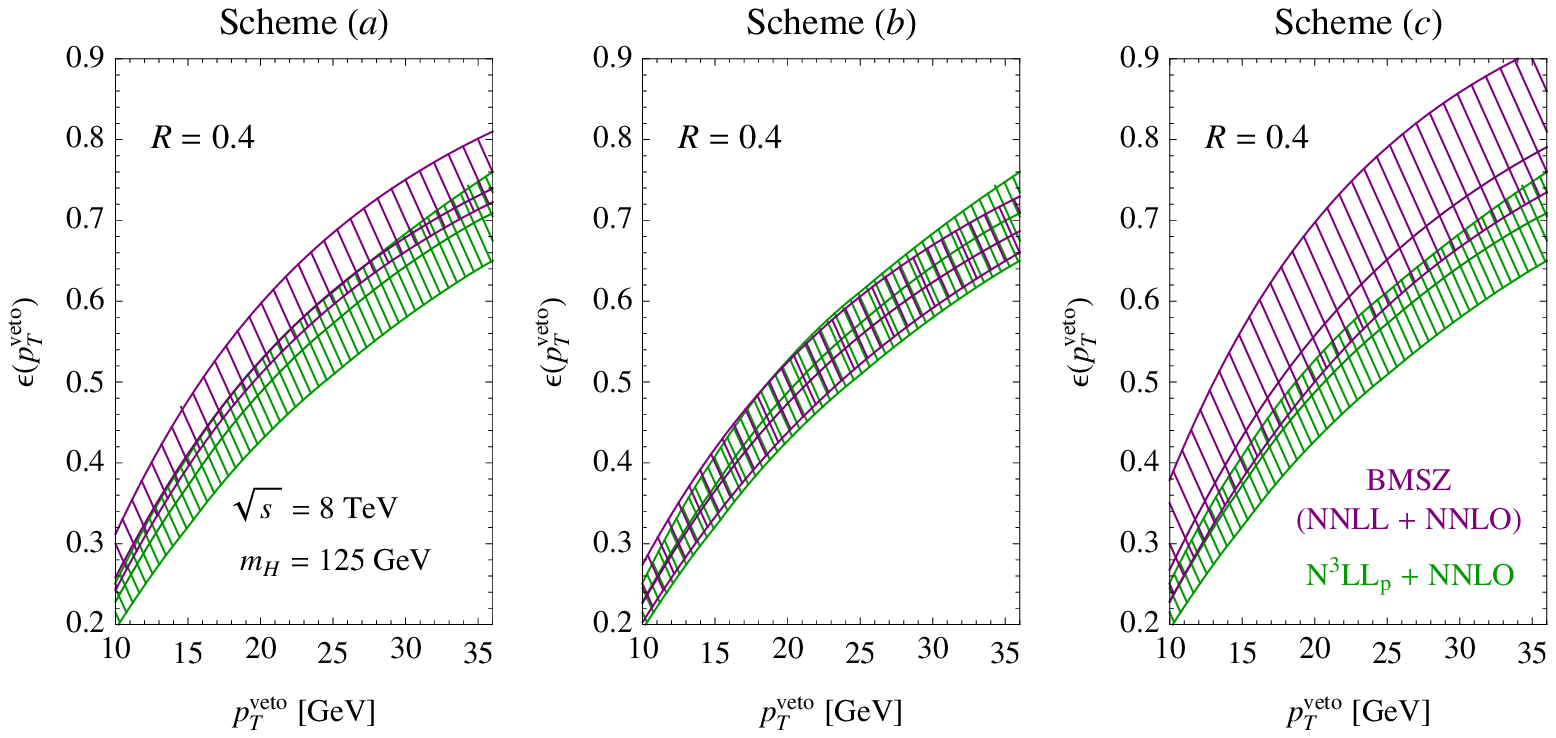}
\vspace{-0.5cm}
\caption{\label{fig:BMSZResults}
Comparison of our result for the jet-veto efficiency (green) to the results of BMSZ (purple) obtained in the three different matching schemes used in \cite{Banfi:2012jm}.}
\end{center}
\end{figure}

In Figure~\ref{fig:BMSZResults}, we compare our findings to the results obtained by Banfi et al.\ (BMSZ) in \cite{Banfi:2012jm}. Their results have NNLL$+$NNLO accuracy and are available through a public code {\sc JetVHeto} \cite{JetVHeto}, which was used to produce the purple bands shown in the figure. An important difference to our approach is that BMSZ consider the efficiency $\epsilon(\pTveto)=\sigma(\pTveto)/\sigma_{\rm tot}$ instead of the cross section itself. Their goal was to divide out the large corrections affecting both $\sigma(\pTveto)$ and the total cross section. We will explain below why we prefer not to follow the same strategy. The perturbative expansion of $\epsilon(\pTveto)$ is not unique. BMSZ consider in their paper three schemes, which translate into three different schemes for how to include the matching corrections. One can either first expand the numerator and the denominator in the formula for $\epsilon(\pTveto)$ and then take their ratio (default scheme $(a)$), expand $\epsilon(\pTveto)$ itself (scheme $(c)$), or consider $1-\epsilon(\pTveto)$ and separately expand numerator and denominator for this quantity (scheme $(b)$). The purple bands in Figure~\ref{fig:BMSZResults} show the scale uncertainty of the BMSZ results, which they obtain by first varying $\mu_f$ and $\mu_r$ by a factor 2 about the default value $m_H/2$, while keeping $1/2<\mu_f/\mu_r<2$ and the resummation scale $Q$ at its default value, and then varying the resummation scale $Q$, while keeping $\mu_f$ and $\mu_r$ fixed at their default values. The bands shown in the figure are the envelope of these variations. 

The difference between the three matching schemes shown in Figure~\ref{fig:BMSZResults} is not negligible. Since the fixed-order corrections to both $\sigma_{\rm tot}$ and $\sigma(\pTveto)$ are large, the different ways of defining the efficiency $\epsilon(\pTveto)$ lead to fairly different results, despite the fact that this difference is formally of ${\cal O}(\alpha_s^3)$. Note that only the virtual part of the corrections cancel in the efficiency $\epsilon(\pTveto)$, since the real-emission corrections to the two cross sections are obviously quite different. The virtual corrections encoded in $C_S(-m_H^2,\mu)$ are indeed responsible for the bad perturbative behavior of the cross section, and they can be avoided by choosing a time-like value $\mu^2=-m_H^2$ for the matching scale \cite{Ahrens:2008qu,Ahrens:2008nc}, as we do in our analysis. By now the virtual corrections to Higgs production are known to three-loop accuracy \cite{Baikov:2009bg,Lee:2010cga,Gehrmann:2010tu}, and the result confirms that the higher-order corrections to $|C_S(-m_H^2,\mu)|^2$ are negligibly small for a time-like scale choice. Even for the standard choice $\mu^2=+m_H^2$, the three-loop corrections are only about 4\%. The part which suffers from these large corrections is thus known very precisely, with sub-percent accuracy. The uncertainty on the fixed-order total cross section is larger, of order 10\%, because the real-emission corrections are not as well known as the virtual part. Dividing by the total cross section therefore increases the uncertainty on the prediction and should better be avoided. 

To compare our results to those of BMSZ, we have divided our prediction for $\sigma(\pTveto)$ by the central value of the resummed total cross section $\sigma_{\rm tot}=19.66_{-0.8\%-7.5\%}^{+2.8\%+7.8\%}$\,pb obtained in \cite{Ahrens:2010rs}, which is a state-of-the-art calculation using the same resummed expression for $C_S(-m_H^2,\mu)$ as we do. The first uncertainty is due to scale variations, whereas the second one is the 90\% C.L.\ error due to the combined PDF and $\alpha_s$ variations. For comparison, we note that the {\sl LHC Higgs Cross Section Working Group\/} adopts the value $\sigma_{\rm tot}=19.52^{+7.2\%+7.5\%}_{-7.8\%-6.9\%}$\,pb \cite{deFlorian:2012yg}. Our results are shown by the green bands in Figure~\ref{fig:BMSZResults}. Note that we do not include an additional uncertainty from the errors on the total cross section, because in order to compare with an experimental cross-section measurement we would have to multiply the efficiency with this same value of the total cross section. We observe that our results are numerically quite similar to the BMSZ results obtained in scheme $(b)$, even though conceptually scheme $(a)$ is closer to our procedure. Note that for their final prediction BMSZ use scheme $(a)$ and its scale variation as the default. As an additional uncertainty, they consider the difference between the default values obtained in the three schemes. As a result, their final uncertainty band is given by the upper edge of the band in scheme $(a)$ and the central curve in scheme $(b)$. The uncertainty band so derived is smaller than the envelope of the bands obtained in the three schemes. 

\begin{table}[t!]
\begin{center}
\begin{tabular}{cccccc}\toprule
& \multicolumn{2}{c}{$R=0.4$}&  & \multicolumn{2}{c}{$R=0.8$} \\ 
\cmidrule{2-3}\cmidrule{5-6}
$ p_T^{\mathrm{veto}}$ [GeV] & $\sigma\left( p_T^{\mathrm{veto}}\right)$ [pb] & $\epsilon\left(p_T^{\mathrm{veto}}\right)$ & &  $\sigma\left( p_T^{\mathrm{veto}}\right)$ [pb] & $\epsilon\left(p_T^{\mathrm{veto}}\right)$ \\
 \midrule
 10 & $\phantom{0}4.48^{ +0.46\,(+0.37)}_{-0.67\,(-0.48)}$  & $0.228^{ +0.023\,(+0.019)}_{-0.034\,(-0.024)}$ & &  $\phantom{0}3.71^{ +0.21\,(+0.19)}_{-0.35\,(-0.34)}$  & $0.189^{ +0.011\,(+0.010)}_{-0.018\,(-0.017)}$ \\
 15 & $\phantom{0}7.31^{ +0.72\,(+0.63)}_{-1.00\,(-0.85)}$  & $0.371^{ +0.036\,(+0.031)}_{-0.051\,(-0.043)}$ & &   $\phantom{0}6.44^{ +0.30\,(+0.28)}_{-0.61\,(-0.59)}$  & $0.328^{ +0.015\,(+0.014)}_{-0.031\,(-0.030)}$  \\
 20 & $\phantom{0}9.57^{ +0.78\,(+0.66)}_{-1.18\,(+1.07)}$& $0.487^{ +0.040\,(+0.034)}_{-0.060\,(-0.055)}$ & &   $\phantom{0}8.71^{ +0.25\,(+0.21)}_{-0.69\,(-0.67)}$  & $0.443^{ +0.013\,(+0.011)}_{-0.035\,(-0.034)}$ \\
 25 & $11.25^{ +0.77\,(+0.65)}_{-1.25\,(-1.15)}$  & $0.572^{ +0.039\,(+0.033)}_{-0.063\,(-0.059)}$ & &   $10.43^{+0.19\,(+0.13)}_{-0.64\,(-0.62)}$  & $0.531^{ +0.010\,(+0.007)}_{-0.033\,(-0.032)}$  \\
 30 & $12.64^{ +0.80\,(+0.67)}_{-1.25\,(-1.15)}$ & $0.643^{ +0.040\,(+0.034)}_{-0.063\,(-0.059)}$& &    $11.86^{+0.18\,(+0.10)}_{-0.57\,(-0.55)}$  & $0.603^{ +0.009\,(+0.005)}_{-0.029\,(-0.028)}$  \\
 35 & $13.75^{ +0.94\,(+0.84)}_{-1.18\,(-1.08)}$ & $0.700^{ +0.048\,(+0.043)}_{-0.060\,(-0.055)}$& &   $13.00^{+0.23\,(+0.18)}_{-0.46\,(-0.43)}$  & $0.662^{ +0.012\,(+0.009)}_{-0.024\,(-0.022)}$ \\
\bottomrule
\end{tabular}
\end{center}
\caption{\label{tab:numbers} 
Numerical results for the jet-veto cross section and efficiency. The uncertainty is obtained by varying $\pTveto/2<\mu<2\pTveto$ and the coefficient $d_3^{\rm veto}(R)$ according to the estimate (\ref{d3estimate}). The numbers in brackets are obtained if only $\mu$ is varied.}
\end{table}

Numerical results for the jet-veto cross section and efficiency are given in Table~\ref{tab:numbers} for two different values of $R$. In addition to the uncertainties shown in the table, one has to account for the PDF and $\alpha_s$ errors. The combined relative uncertainty due to these input parameters is very similar to that obtained for the total cross section. The table shows that it would be beneficial to increase the jet radius compared to the presently used values $R\approx 0.4$. On the experimental side, this would increase the sensitivity to the underlying event and pile-up, but these effects could be mitigated by resorting to techniques such as the one proposed in \cite{Soyez:2012hv}.

\section{Conclusion}
\label{sec:concl}

Using methods from effective field theory, we have obtained precise predictions for the Higgs-boson production cross section in the presence of a jet veto, in which both Sudakov logarithms of the form $\alpha_s^n\ln^m(\pTveto/m_H)$ arising due to the veto as well as the large virtual corrections affecting also the total cross section are resummed in a systematic way. To demonstrate the validity of the factorization formula (\ref{sigfinal}) for the cross section, which was first derived in \cite{Becher:2012qa}, we have determined all its ingredients at two-loop order and have increased the logarithmic accuracy of the resummation to (partial) N$^3$LL order. In particular, we have computed the two-loop anomaly coefficient $d_2^{\rm veto}(R)$ and have obtained a fully analytic expression for its series expansion valid for $R<\pi$, including the constant term, which previously was only known in numerical form. Our result agrees with the expression obtained in \cite{Banfi:2012yh} through a calculation in QCD. Contrary to claims in the literature \cite{Tackmann:2012bt}, we find that even for $R={\cal O}(1)$ soft-collinear mixing contributions, which would break factorization, are absent. This establishes factorization at NNLL accuracy. 

In addition to the explicit two-loop calculation, we have discussed in detail why such soft-collinear mixing terms are absent also in higher orders of the perturbative expansion. That they do not arise becomes manifest if the multipole expansion in the effective theory is properly implemented, i.e., if power-suppressed terms are consistently expanded away at the integrand level. Since SCET does not include hard cutoffs to separate the soft and collinear momentum regions, this multipole expansion is necessary in order to avoid double counting of certain momentum configurations. If, as in \cite{Tackmann:2012bt}, the multipole expansion is only performed at the Lagrangian level, but not for the measurement function which defines the observable, then soft-collinear mixing terms arise in the overlap of the soft and collinear regions. However, we have shown that they cancel against the contribution from the overlap region. Our analysis thus reinforces the validity of the factorization theorem proposed in \cite{Becher:2012qa}.

We have extended the phenomenological analysis of the Higgs-boson production cross section with a jet veto to N$^3$LL$_{\rm p}$ accuracy, where the subscript ``p'' stands for ``partial'' and indicates that two perturbative coefficients -- the four-loop cusp anomalous dimension and the three-loop coefficient of the collinear anomaly -- are currently still missing to complete the resummation at this order. The motivation for going beyond NNLL order is that the two-loop anomaly coefficient $d_2^{\rm veto}(R)$ turns out to be numerically quite large. For example, at $R=0.4$ it is almost six times larger than the corresponding coefficient arising in the resummation formula for the transverse-momentum spectrum of the Higgs boson. This gives rise to a significant scale uncertainty of the NNLL result. Furthermore, for small values of the jet-radius parameter $R$, the scale variation cannot be trusted as an estimator of the effect of higher-order corrections. This is because the three-loop coefficient $d_3^{\rm veto}(R)$ is enhanced by two powers of $\ln R$, while the scale-dependent pieces at two-loop order involve at most a single logarithm of $R$. In view of this fact, we have estimated the impact of the three-loop coefficient and find that it is relatively small, even with a generous estimate for the coefficient of the double-logarithmic term. The numerical impact of the other missing ingredient, the four-loop cusp anomalous dimension, is negligibly small. The most important contribution at N$^3$LL order arises from the two-loop beam functions, which we have extracted numerically using the fixed-order code {\sc HNNLO}. As for the two-loop anomaly coefficient, we find that the two-loop perturbative corrections to the beam functions are rather large, especially in the region of small $R$, where they are logarithmically enhanced. For $R<0.8$, the two-loop corrections are larger than estimated by the one-loop scale variation, and also the scale uncertainty of the two-loop beam functions is larger than at one-loop order. It constitutes the main uncertainty of our final results for the cross section. On the other hand, since the leading $\ln R$-dependent corrections to both the exponent of the collinear anomaly and the beam functions are included in our N$^3$LL$_{\rm p}$+NNLO results, we expect that yet higher-order corrections would turn out to be small. Nevertheless, it would be interesting and important to compute the $\ln R$-enhanced terms at three-loop order (and perhaps even beyond). We believe that it should be possible to calculate the leading logarithmic part of the three-loop anomaly coefficient $d_3^{\rm veto}(R)$, which would reduce the uncertainty of our predictions especially at low values of $R$. 

In addition to performing the resummation of the jet-veto cross section at leading power in $\pTveto/m_H$, we have matched our resummed results to the full fixed-order expression for the cross section computed at NNLO, which has allowed us to also include power-suppressed terms. The size of these power corrections serves as an important check of the quality of the expansion in the small ratio $\pTveto/m_H$, which is the basis for our factorization formula. Numerically, the power corrections turn out to be quite small for the relevant values of $\pTveto$. Importantly, we find no evidence for a growth of the power corrections with increasing $R$. A crucial element of our analysis is that we have separated different sources of theoretical uncertainties. This avoids accidental cancellations and furthermore allows us to investigate the uncertainty associated with each source individually. In our final results, we have added in quadrature the uncertainties in the hard function (which contains the resummation of all large logarithms), the beam functions, and the power corrections to the cross section. 
 
Our results significantly improve the accuracy of the Higgs-boson production cross section with a jet veto. Detailed numerical predictions can be found in Table~\ref{tab:numbers}. It would be interesting to perform the resummation of large logarithms also for the $W^+ W^-$ production cross section, which is the main background to the $H\to W^+ W^-$ signal in the presence of a jet veto. The $W^+ W^-$ cross section is furthermore used to search for anomalous gauge couplings, and also in this case it is necessary to impose a jet veto. Likewise, Sudakov logarithms should also be resummed for the cross sections for Higgs production in association with one or two tagged jets. The resummation is more challenging in the higher jet bins, due to the presence of non-global logarithms. However, a recent study at NLL order suggests that the numerical effect of the non-global logarithms is likely to be small \cite{Liu:2012sz,Liu:2013hba}. 

{\em Acknowledgments:\/} 
We are grateful to Andrea Banfi, Guido Bell, Gavin Salam, Frank Tackmann, and Giulia Zanderighi for useful discussions. T.B.\ and M.N.\ thank KITP Santa Barbara for hospitality and support during the completion of this paper. KITP is supported by the National Science Foundation under Grant No.\ NSF PHY11-25915. The work of T.B.\ is supported by the Swiss National Science Foundation (SNF) under grant 200020-140978. The research of M.N.\ is supported by the ERC Advanced Grant EFT4LHC, the Cluster of Excellence {\em Precision Physics, Fundamental Interactions and Structure of Matter\/} (PRISMA -- EXC 1098) and grant NE 398/3-1 of the German Research Foundation (DFG), grants 05H09UME and 05H12UME of the German Federal Ministry for Education and Research (BMBF), and the Rhineland-Palatinate Research Center Elementary Forces and Mathematical Foundations.

\newpage
\begin{appendix}
\numberwithin{equation}{section}

\section{\boldmath Series expansion of $f_B(R)$ for $R<\pi$}
\label{App:correlated}

In this appendix, we describe the analytical calculation of the function $f_B(R)$, which enters the two-loop anomaly coefficient in (\ref{d2vetoR}), as a series expansion valid for $R<\pi$. Since the contribution from two independent gluon emissions, given by the second term in (\ref{fBRres}), is known in closed form, we will focus on the correlated emission contribution (\ref{softcorr}). The corresponding squared matrix elements ${\cal A}_s(k,l)$ can be found in compact form in Appendix~C of \cite{Becher:2012qc} and also in \cite{Tackmann:2012bt}. Using relations (\ref{Deltad2}), (\ref{Deltadd2}) and (\ref{softcorr}), we find that (recall that $C_B=C_A$ for the Higgs case)
\begin{equation}
   f_B(R) = C_A I_A + T_F n_f I_f + C_B \left( - \frac{\pi^2 R^2}{12} + \frac{R^4}{16} \right) ,
\end{equation}
where
\begin{equation}\label{Ifdef}
   I_f = 4\int_0^{1/2}\!dz \int_0^\infty\!d\Delta y \int_0^\pi\!\frac{d\Delta\phi}{\pi}\,
    \theta(\Delta y^2+\Delta\phi^2-R^2)\,A_f(z,\Delta y,\Delta\phi) \,,
\end{equation}
and similarly for the integral $I_{\rm sum}\equiv I_A+\frac12\,I_f$. The reduced matrix elements $A_i$ are given by
\begin{equation}\label{startform}
\begin{aligned}
   A_f(z,\Delta y,\Delta\phi) &= \frac14 \ln\left[ \frac{\sqrt{1-4z(1-z)\sin^2\frac{\Delta\phi}{2}}}{1-z} \right]
    \frac{1}{\sinh^2\frac{\Delta y}{2}+\sin^2\frac{\Delta\phi}{2}}\,
    \frac{1}{1+4z(1-z)\sinh^2\frac{\Delta y}{2}} \\
   &\quad\times \left[ 1 - \frac{4z(1-z)\sinh^2\frac{\Delta y}{2}\cosh^2\frac{\Delta y}{2}}%
    {(\sinh^2\frac{\Delta y}{2}+\sin^2\frac{\Delta\phi}{2})\,(1+4z(1-z)\sinh^2\frac{\Delta y}{2})} 
    \right] , \\[2mm]
   A_{\rm sum}(z,\Delta y,\Delta\phi) &= \frac14 \ln\left[ \frac{\sqrt{1-4z(1-z)\sin^2\frac{\Delta\phi}{2}}}{1-z} 
    \right] \frac{1}{\sinh^2\frac{\Delta y}{2}+\sin^2\frac{\Delta\phi}{2}}\,
    \frac{1}{1+4z(1-z)\sinh^2\frac{\Delta y}{2}} \\
   &\quad\times \left[ \bigg( \frac{z^2+(1-z)^2}{z(1-z)} + 1 + 2\sinh^2\frac{\Delta y}{2} \bigg)
    \bigg( 1 - 2\sin^2\frac{\Delta\phi}{2} \bigg) - \frac12 \right] .
\end{aligned}
\end{equation}
We now change variables to $\xi=4z(1-z)$, $u=\sin\frac{\Delta\phi}{2}$, and $v=\sinh\frac{\Delta y}{2}$. This yields
\begin{equation}\label{If}
\begin{aligned}
   I_f &= \frac{1}{2\pi} \int_0^1\!\frac{d\xi}{\sqrt{1-\xi}} \int_0^\infty\!\frac{dv}{\sqrt{1+v^2}}
    \int_0^1\!\frac{du}{\sqrt{1-u^2}}\,
    \ln\left[\frac{4(1-\xi u^2)}{(1+\sqrt{1-\xi})^2} \right] 
    \frac{1}{u^2+v^2}\,\frac{1}{1+\xi v^2} \\
   &\quad\times \left[ 1 - \frac{\xi v^2(1+v^2)}{(u^2+v^2)\,(1+\xi v^2)} \right] 
    \theta\bigg(\! \arcsin^2u + \mbox{arcsinh}^2v - \frac{R^2}{4} \bigg) \,, \\[2mm]
   I_{\rm sum} &= \frac{1}{2\pi} \int_0^1\!\frac{d\xi}{\sqrt{1-\xi}}
    \int_0^\infty\!\frac{dv}{\sqrt{1+v^2}} \int_0^1\!\frac{du}{\sqrt{1-u^2}}\,
    \ln\left[\frac{4(1-\xi u^2)}{(1+\sqrt{1-\xi})^2} \right] 
    \frac{1}{u^2+v^2}\,\frac{1}{1+\xi v^2} \\
   &\quad\times \left[ \frac{2(2-\xi)}{\xi}\,(1-2u^2) - 2u^2 + 2v^2 - 4u^2 v^2 + \frac12 \right] 
    \theta\bigg(\! \arcsin^2u + \mbox{arcsinh}^2v - \frac{R^2}{4} \bigg) \,. 
\end{aligned}
\end{equation}

\subsection{\boldmath Asymptotic behavior for $R\to 0$}

For very small $R$, the main sensitivity to the jet-radius parameter arises from the region where both $u$ and $v$ are very small. Expanding the integrands about this limit, we obtain the simpler expressions
\begin{equation}
\begin{aligned}
   I_f^{\rm exp} &= \frac{1}{\pi} \int_0^1\!\frac{d\xi}{\sqrt{1-\xi}}\,
    \ln\left[\frac{2}{1+\sqrt{1-\xi}} \right] \int_0^\infty\!dv 
    \int_0^1\!du\,\frac{1}{u^2+v^2} \left[ 1 - \frac{\xi v^2}{u^2+v^2} \right] 
    \theta\bigg(\! u^2 + v^2 - \frac{R^2}{4} \bigg) \,, \\[2mm]
   I_{\rm sum}^{\rm exp} &= \frac{1}{\pi} \int_0^1\!\frac{d\xi}{\sqrt{1-\xi}}\,
    \ln\left[\frac{2}{1+\sqrt{1-\xi}} \right] \int_0^\infty\!dv 
    \int_0^1\!du\,\frac{1}{u^2+v^2} \left[ \frac{4}{\xi} - \frac32  \right] 
    \theta\bigg(\! u^2 + v^2 - \frac{R^2}{4} \bigg) \,.
\end{aligned}
\end{equation}
The double integrals over $u$ and $v$ can be evaluated using polar coordinates $u=r\sin\varphi$ and $v=r\cos\varphi$, carrying out the integration over $r$ first. Performing then the integrations over $\varphi$ and $\xi$, we find
\begin{equation}\label{expandedints}
\begin{aligned}
   I_f^{\rm exp} &= \left( \frac23\ln 2 - \frac{23}{36} \right) \ln R 
    - \frac{13}{72} + \frac{13}{9} \ln 2 - \frac43 \ln^2 2 \,, \\
   I_{\rm sum}^{\rm exp} &= \left( \frac32 - \frac{\pi^2}{6} - \frac32\ln 2 \right) 
    \big( \ln R - 2\ln 2 \big) \,.
\end{aligned}
\end{equation}

The differences $\Delta I_i=(I_i-I_i^{\rm exp})$ can be evaluated by setting $R=0$. We obtain
\begin{equation}\label{DeltaIf}
\begin{aligned}
   \Delta I_f
   &= \frac{1}{2\pi} \int_0^1\!\frac{d\xi}{\sqrt{1-\xi}} \int_0^\infty\!dv \int_0^1\!du\,
    \frac{1}{u^2+v^2} \\
   &\quad\times \Bigg\{ \frac{1}{\sqrt{1+v^2}}\,\frac{1}{\sqrt{1-u^2}}\,
    \ln\left[\frac{4(1-\xi u^2)}{(1+\sqrt{1-\xi})^2} \right] \frac{1}{1+\xi v^2} 
    \left[ 1 - \frac{\xi v^2(1+v^2)}{(u^2+v^2)\,(1+\xi v^2)} \right] \\
   &\hspace{1.2cm}\mbox{}- \ln\left[\frac{4}{(1+\sqrt{1-\xi})^2} \right] 
    \left[ 1 - \frac{\xi v^2}{u^2+v^2} \right] \Bigg\} \,,
\end{aligned}
\end{equation}
and similarly for $\Delta I_{\rm sum}$. In the next step we perform the integral over $v$. In the first double integral we perform a partial fraction decomposition and then use the integrals
\begin{equation}\label{vints}
\begin{aligned}
   I_1(u) &= \int_0^\infty\!\frac{dv}{\sqrt{1+v^2}}\,\frac{1}{u^2+v^2}
    = \frac{\arccos u}{u\sqrt{1-u^2}} \,, \\
   I_2(\xi) &= \int_0^\infty\!\frac{dv}{\sqrt{1+v^2}}\,\frac{1}{1+\xi v^2}
    = \frac{1}{2\sqrt{1-\xi}}\,\ln\frac{1+\sqrt{1-\xi}}{1-\sqrt{1-\xi}} \,.  
\end{aligned}
\end{equation}
Integrals with squared denominators can be obtained from these expressions by means of derivatives. After the integration over $v$ has been performed, the contribution resulting from the terms in the second line of (\ref{DeltaIf}) exhibits a pole at $u=0$, which is cancelled by the integral over the subtraction term shown in the third line. This cancellation can be made explicit using the identity
\begin{equation}\label{amazing}
   \int_0^1\!du \left[ \frac{\arccos u}{u(1-u^2)} - \frac{\pi}{2u} \right] = 0 \,.
\end{equation}
This gives
{\small
\begin{equation}\label{DIf}
\begin{aligned}
   \Delta I_f
   &= \frac{1}{2\pi} \int_0^1\!\frac{d\xi}{\sqrt{1-\xi}} \ln\frac{2}{1+\sqrt{1-\xi}} 
    \int_0^1\!\frac{du}{\sqrt{1-u^2}}\,\Bigg\{ \frac{2\xi}{(1-\xi u^2)^2} \\
   &\quad\mbox{}+ \left[ \frac{2-\xi-u^2\xi(2+3\xi)+4u^4\xi^2}{(1-\xi u^2)^3} - (2-\xi)
    \right] I_1(u) 
    - \frac{\xi\big[ 4-3\xi-u^2\xi(2+\xi)+2u^4\xi^2 \big]}{(1-\xi u^2)^3}\,I_2(\xi) \Bigg\} \\
   &\quad\mbox{}+ \frac{1}{2\pi} \int_0^1\!\frac{d\xi}{\sqrt{1-\xi}}
    \int_0^1\!\frac{du}{\sqrt{1-u^2}} \ln(1-\xi u^2) \\
   &\quad\times \Bigg\{ \frac{\xi}{(1-\xi u^2)^2}
    + \frac{2-\xi-u^2\xi(2+3\xi)+4u^4\xi^2}{2(1-\xi u^2)^3}\,I_1(u) 
    - \frac{\xi\big[ 4-3\xi-u^2\xi(2+\xi)+2u^4\xi^2 \big]}{2(1-\xi u^2)^3}\,I_2(\xi) \Bigg\} ,
\end{aligned}
\end{equation}}
and similarly
\begin{equation}\label{DIsum}
\begin{aligned}
   \Delta I_{\rm sum}
   &= \frac{1}{2\pi} \int_0^1\!\frac{d\xi}{\sqrt{1-\xi}} \ln\frac{2}{1+\sqrt{1-\xi}} 
    \int_0^1\!\frac{du}{\sqrt{1-u^2}} \\
   &\quad\times \Bigg\{ 
    \left[ \frac{8-3\xi-16u^2+8\xi u^4}{\xi(1-\xi u^2)} - \frac{8-3\xi}{\xi} \right] I_1(u) 
    - \frac{4-3\xi-4u^2(2-\xi)}{(1-\xi u^2)}\,I_2(\xi) \Bigg\} \\
   &\quad\mbox{}+ \frac{1}{2\pi} \int_0^1\!\frac{d\xi}{\sqrt{1-\xi}}
    \int_0^1\!\frac{du}{\sqrt{1-u^2}} \ln(1-\xi u^2) \\
   &\quad\times \Bigg\{ \left[ \frac{8-3\xi-16u^2+8\xi u^4}{2\xi(1-\xi u^2)} \right] I_1(u) 
    - \frac{4-3\xi-4u^2(2-\xi)}{2(1-\xi u^2)}\,I_2(\xi) \Bigg\} \,.
\end{aligned}
\end{equation}

In the next step we integrate over $u$. This is straightforward for all terms except those involving the function $I_1(u)$ in the second contribution in expressions (\ref{DIf}) and (\ref{DIsum}). For these terms, one needs the basis integrals
\begin{equation}\label{Kndef}
\begin{aligned}
   J(\xi) &= \frac{1}{4\pi} \int_0^1\!dt\,\ln(1-\xi t)\,\frac{\arccos\sqrt{t}}{t} \,, \\
   K_n(\xi) &= \frac{1}{4\pi} \int_0^1\!dt\,
    \frac{\ln(1-\xi t)}{\left(1-\xi t\right)^n}\,\frac{\arccos\sqrt{t}}{t(1-t)} \,;
    \quad n=0,1,2,3 \,,
\end{aligned}
\end{equation}
where we have changed variables from $u$ to $t=u^2$. Rewriting 
\begin{equation}
   \ln(1-\xi t) = -t \int_0^\xi\!dy\,\frac{1}{1-y t} \,,
\end{equation}
we first perform the integral over $t$ and then integrate over $y$. In this way, we obtain
\begin{equation}
\begin{aligned}
   J(\xi) &= \frac14\,L_2\bigg( - \frac{1-x}{1+x} \bigg) + \frac14\,\ln^2\frac{1+x}{2} \,, \\
   K_0(\xi) &= - \frac12\,L_2(-x) - \frac{\pi^2}{24} \,, \\
   K_1(\xi) &= \frac14 \left( \frac{1}{x^2} - 1 \right) L_2\bigg( \frac{1-x}{1+x} \bigg)
    + \frac{K_0(\xi)}{x^2} \,, \\
   K_2(\xi) &= \frac{1-x^2}{4x^3} \left[ \ln\frac{1+x}{2x} - \frac{1-x}{2} \right]
    + \frac14 \left( \frac{1}{x^4} - 1 \right) L_2\bigg( \frac{1-x}{1+x} \bigg) 
    + \frac{K_0(\xi)}{x^4} \,, \\  
   K_3(\xi) &= \frac{2+3x^2-5x^4}{32 x^4}
    + \frac{5(1-x^4)}{16x^5} \left[ \ln\frac{1+x}{2x} - \frac{7}{20} \right]
    + \frac14 \left( \frac{1}{x^6} - 1 \right) L_2\bigg( \frac{1-x}{1+x} \bigg)
    + \frac{K_0(\xi)}{x^6} \,,
\end{aligned}
\end{equation}
where $x=\sqrt{1-\xi}$. Using these results, the integrations over $u$ can be performed in a straightforward way. Integrating the resulting expressions over $\xi$, which is facilitated by changing variables from $\xi$ to $x$, we obtain
\begin{equation}
\begin{aligned}
   \Delta I_f &= \frac{353}{216} - \frac{\pi^2}{18} - \frac73 \ln 2 + \frac23 \ln^2 2 \,, \\
   \Delta I_{\rm sum} &= - 3 + \frac{\pi^2}{8} + \frac92 \ln 2 - \frac{\pi^2}{3} \ln 2
    - \frac32 \ln^2 2 + \frac{\zeta_3}{2} \,.
\end{aligned}   
\end{equation}

In a last step, we now add back the subtraction terms given in (\ref{expandedints}). Reexpressing the combination $I_{\rm sum}$ in terms of $I_f$ and $I_A$, we finally arrive at
\begin{equation}\label{Iifinal}
\begin{aligned}
   I_f &= \left( \frac23\ln 2 - \frac{23}{36} \right) \ln R 
    + \frac{157}{108} - \frac{\pi^2}{18} - \frac89 \ln 2 - \frac23 \ln^2 2 
    + {\cal O}(R^2) \,, \\
   I_A &= \left( \frac{131}{72} - \frac{\pi^2}{6} - \frac{11}{6} \ln 2 \right) \ln R 
    - \frac{805}{216} + \frac{11\pi^2}{72} + \frac{35}{18} \ln 2 + \frac{11}{6} \ln^2 2 
    + \frac{\zeta_3}{2} + {\cal O}(R^2) \,.
\end{aligned}
\end{equation}
From these results, one derives the coefficients $c_L^i$ and $c_0^i$ given in (\ref{cLtoc4}).

\subsection{\boldmath Series expansion in powers of $R^2$}

Except for the leading logarithmic singularity exhibited in (\ref{Iifinal}), the integrals $I_f$ and $I_A$ can be expanded in a power series in $R^2$, provided that $R<\pi$. Terms of ${\cal O}(R^{2n})$ with $n\ge 1$ can be obtained by applying $n$ differential operators $\partial_{R^2}$ on the subtracted integrals $\Delta I_i$. To this end, we introduce polar coordinates $\Delta\phi=r\sin\varphi$ and $\Delta y=r\cos\varphi$ and rewrite (\ref{Ifdef}) in the form (with $i=f,A$)
\begin{equation}
   \Delta I_i = \frac{2}{\pi} \int_0^{1/2}\!dz \int_0^{\pi/2}\!d\varphi
    \int_0^{\pi^2/\sin^2\varphi}\!\!dr^2\,\theta(r^2-R^2)\,
    \Delta A_i(z,r\cos\varphi,r\sin\varphi) 
   = \mbox{const.} + \delta I_i(R^2) \,,
\end{equation}
where $\delta I_i(R^2)$ is a power series in $R^2$, and the constant term will be irrelevant for our discussion. The quantities $\Delta A_i$ denote the reduced matrix elements $A_i$ in (\ref{startform}) with their leading singularities (for $\Delta\phi,\Delta y\to 0$) subtracted. The first derivative $\partial_{R^2}$ yields a $\delta$-distribution, $\partial_{R^2}\,\theta(r^2-R^2)=-\delta(r^2-R^2)$, and for $R<\pi$ the radial integral then simply sets $r=R$ in the integrand. It is then straightforward to show that
\begin{equation}\label{c2ni}
   \delta I_i(R^2) = \sum_{n=1}^\infty\,c_{2n}^i R^{2n} \,, \quad \mbox{with} ~~
   c_{2n}^i = - \frac{1}{n} \int_0^{1/2}\!dz\,\frac{2}{\pi} \int_0^{\pi/2}\!d\varphi\,
    a_{2n}^i(z,\varphi) \,,
\end{equation}
where the expansion coefficients are defined by
\begin{equation}
   R^2 A_i(z,R\cos\varphi,R\sin\varphi) = \sum_{n=0}^\infty\,a_{2n}^i(z,\varphi)\,R^{2n} \,.
\end{equation}
The remaining integrals over $\varphi$ and $z$ in (\ref{c2ni}) can be performed in closed form. Explicitly, we obtain 
\begin{equation}
\begin{aligned}
   c_2^f &= \frac{3071}{86400} - \frac{7}{360} \ln 2 
    = 0.0220661 \,, \\
   c_4^f &= - \frac{168401}{101606400} + \frac{53}{30240} \ln 2 
    = -0.000442544 \,, \\
   c_6^f &= \frac{7001023}{48771072000} - \frac{11}{100800} \ln 2
    = 0.0000679076 \,, \\
   c_8^f &= - \frac{5664846191}{566524772352000} + \frac{4001}{479001600} \ln 2
    = - 4.20958\cdot 10^{-6} \,, \\
   c_{10}^f &= \frac{68089272001}{83774850711552000} - \frac{13817}{21794572800} \ln 2
    = 3.73334\cdot 10^{-7} \,,
\end{aligned}
\end{equation}
and
\begin{equation}
\begin{aligned}
   c_2^A &= \frac{1429}{172800} + \frac{\pi^2}{48} + \frac{13}{180} \ln 2 
    = 0.263947 \,, \\
   c_4^A &= - \frac{9383279}{406425600} - \frac{\pi^2}{3456} + \frac{587}{120960} \ln 2 
    = - 0.0225794 \,, \\
   c_6^A &= \frac{74801417}{97542144000} - \frac{23}{67200} \ln 2
    = 0.000529625 \,, \\
   c_8^A &= - \frac{50937246539}{2266099089408000} - \frac{\pi^2}{24883200}
    + \frac{28529}{1916006400} \ln 2
    = - 0.0000125537 \,, \\
   c_{10}^A &= \frac{348989849431}{243708656615424000} - \frac{3509}{3962649600} \ln 2
    = 8.18201\cdot 10^{-7} \,.
\end{aligned}
\end{equation}
Even for a large value such as $R=2$, the two power series converge rapidly, and truncating them at the $R^{10}$ term provides results that are accurate to the few permille level. For $R\le 1$, it suffices to keep the first few terms in the series.

\end{appendix}

\end{document}